\documentclass[twocolumn]{aastex631}
\usepackage{newtxtext,newtxmath}
\usepackage{fix-cm}
\usepackage[T1]{fontenc}
\usepackage{textcomp}
\usepackage{amsmath}
\usepackage{natbib}
\usepackage{multirow} 
\usepackage{gensymb}
\usepackage{longtable}
\usepackage[utf8]{inputenc}
\DeclareUnicodeCharacter{02BC}{'}
\usepackage{placeins}
\usepackage{xcolor}
\usepackage{hyperref}

\begin{document} 

\title{\large{Multimessenger signatures of a deformed magnetar in gamma-ray bursts}}

\author{\textcolor{blue!50!black}{Parisa Hashemi}}

\affiliation{Department of Physics, Isfahan University of Technology, Isfahan 84156-83111, Iran}

\author{\textcolor{blue!50!black}{Soroush Shakeri}}

\affiliation{Department of Physics, Isfahan University of Technology, Isfahan 84156-83111, Iran}
\affiliation{Iranian National Observatory, Institute for Research in Fundamental Sciences (IPM), P. O. Box 19395-5531 Tehran, Iran}
\affiliation{ICRANet-Isfahan, Isfahan University of Technology, Isfahan 84156-83111, Iran}

\author[0000-0001-7959-3387]{\textcolor{blue!50!black}{Yu Wang}}

\affiliation{International Center for Relativistic Astrophysics Network (ICRANet), Pescara I-65122, Italy}
\affiliation{International Center for Relativistic Astrophysics (ICRA), Roma I-00185, Italy}
\affiliation{National Institute for Astrophysics (INAF) -- Osservatorio Astronomico d'Abruzzo, Teramo I-64100, Italy.}

\author{\textcolor{blue!50!black}{Liang Li}}

\affiliation{School of Physical Science and Technology, Ningbo University, Ningbo, Zhejiang 315211, Peopleʼs Republic of China}
\affiliation{Institute of Fundamental Physics and Quantum Technology, Ningbo University, Ningbo, Zhejiang 315211, People’s Republic of China}
\affiliation{National Institute for Astrophysics (INAF) -- Osservatorio Astronomico d'Abruzzo, Teramo I-64100, Italy.}

\author[0000-0002-2516-5894]{\textcolor{blue!50!black}{Rahim Moradi}}

\affiliation{Key Laboratory of Particle Astrophysics, Institute of High Energy Physics, Chinese Academy of Sciences, Beijing
100049, China.}

\begin{abstract}

We study the evolution  of a newly formed magnetized neutron-star (NS)   as a power source  of  gamma-ray bursts (GRBs) in the light of  both gravitational-wave (GW) and electromagnetic (EM) radiations. 
The compressible and incompressible fluids  are employed in order to model the secular evolution of stable Maclaurian spheroids. It is shown that the GW and EM emissions  evolve as a function of eccentricity and rotational frequency with time. We find that the luminosity  characteristics crucially depend  on NS parameters  such as  magnitude and  structure of  magnetic field, ellipticity and  the equation of state (EoS) of the fluid. The presence of X-ray flares, whose origins are not yet well understood, can be captured in our model regarding some specific nuclear EoSs. Our model allowing us to explain flares that occur within the wide range of $ 10$ to $10^4$ s and the peak EM luminosity in the order of $10^{46}$ - $10^{51}$ $\rm \text{erg} s^{-1}$ by using a reasonable set of parameters,  such as magnetic field strength around $10^{14}-10^{16}$ G, the quadrupole-to-dipole ratio of magnetic field up to 500. By applying our model to a sample of GRB X-ray flares observed by the Swift/X-ray Telescope, we try to constraint the crucial parameters of a deformed magnetar via a Marcov Chain Monte Carlo fitting method. Our analysis shows that ongoing and upcoming joint multimessenger detections can be used to understand the nature of a GRB's central engine and its evolution at the early times of the burst formation. 

\end{abstract}


\section{Introduction}

Gamma-ray bursts (GRBs) are broadly classified, based on their observed duration, into short and long bursts \citep{1993ApJ...413L.101K}, each associated with distinct progenitor systems \citep{2004RvMP...76.1143P, 2019pgrb.book.....Z}. Long GRBs, with durations exceeding 2s, are generally linked to the collapse of massive stars, often Wolf–Rayet stars, at the end of their evolution \citep{1998AAS...193.4207M,2001ApJ...550..410M}, and the associated supernova signals are confirmed \citep{1998Natur.395..670G, 2008ApJ...687.1201K, 2012ApJ...759..107K}. Short GRBs, typically lasting less than 2s, are thought to originate from the mergers of binary neutron star (NS) or NS–black hole (BH) systems, driven by gravitational wave (GW) inspirals \citep{2014ARA&A..52...43B, 2015JHEAp...7...73D, 2023A&A...669A..36V}. 

The final outcome of these events—a massive NS (potentially a magnetar) or a BH—depends on factors including the total mass of the remnant, its rotation rate, and the equation of state (EoS) of nuclear matter. The maximum gravitational mass that a non-rotating NS can support, known as the Tolman–Oppenheimer–Volkoff (TOV) limit, is estimated to be $2.2$–$2.4\,M_{\odot}$, though the precise value depends on the EoS of ultradense matter \citep{2012ARNPS..62..485L}. If the remnant mass from a binary NS merger or a massive star collapse largely exceeds the TOV limit, a BH forms promptly  \citep{1998AAS...193.4207M, 2011ApJ...730...70O}. In cases of rapid or differential rotation, however, the remnant may temporarily exist as a hypermassive NS, stabilized by centrifugal forces \citep{2006PhRvL..96c1101D,2016PhRvD..93d4065G}. In certain cases, particularly when the total remnant mass is only slightly above the TOV limit and the magnetic fields are strong, the remnant may persist as a supramassive NS for hours to days \citep{2017PhRvD..95f3016C}. Gradual spin-down reduces its centrifugal support, potentially delaying collapse to a BH or preventing it altogether if the core mass remains below the critical threshold \citep{2013ApJ...771L..26G,2021GReGr..53...59S}.

The ability of an NS to act as the central engine of a GRB depends on its capacity to produce ultrarelativistic, baryon-poor jets. Newly formed NSs, especially those created in core-collapse events, face challenges due to baryon loading \citep{2000ApJ...529..146E}. The high initial temperatures of proto-NSs ($T \sim 10^{10} \, \mathrm{K}$) generate intense neutrino fluxes that ablate baryonic material from the NS surface, increasing the jet’s baryonic mass and inhibiting the formation of high Lorentz factor jets \citep{2017ApJ...846..114F}. In contrast, BH-powered jets are generally cleaner because energy extraction through mechanisms such as the Blandford–Znajek process occurs in the baryon-poor polar regions \citep{2000PhR...325...83L}. One possible solution involves strong magnetic fields ($B > 10^{14} \, \mathrm{G}$) of magnetars, which can confine the jet to narrow, baryon-poor polar regions, thereby reducing baryonic contamination \citep{2007MNRAS.380.1541B,2021MNRAS.508.5390S,2024MNRAS.528.3705W}. Additionally, the observed jet may be produced after the baryons in the surrounding environment have been cleared. Once the jet is formed, it is initially responsible for opening the enclosing shells by creating a cocoon structure. During this process, the initial jet clears the baryonic material along its path, allowing a late, clean, relativistic jet core to emerge \citep{2005ApJ...629..903L,2011MNRAS.413.2031M}.


Observationally, there are several pieces of evidence showing that the GRB and its afterglow may be powered by a rapidly rotating, highly magnetized NS in which loss of angular momentum is a result of EM and GW emissions or is due to the accretion processes \citep{2013MNRAS.430.1061R,Mosta:2020hlh,Bucciantini:2012sf,Bianco:2023nvk,sarin2}. To date, GW 170817/GRB 170817A \citep{2017ApJ...848L..12A} remains the only observed GW event from a binary NS merger, with multiple perspectives on the nature of its outcome. For example, \citet{2023A&A...669A..36V} argued that a Kerr BH was ultimately formed, inferred from the descending GW chirp that exceeds the limits of a NS, with a 1.7-second delay between the GW signal and EM emission attributed to the collapse of a hypermassive NS into a BH. \citet{2019ApJ...876..139G} suggested that the remnant collapsed into a BH after emitting blue and red kilonova ejecta. The timeline of the collapse aligns with the delay needed for the relativistic jet to emerge. \citet{2018ApJ...861..114Y} propose that a long-lived NS could have formed post-merger, which could explain late-time kilonova emission due to energy injection. The  plateau phase in X-ray afterglow emission is observed in many GRBs  \citep[e.g.,][ and references therein]{Liang2007,Li2012,Li2015,Dainotti2020,Dainotti2022} and can be explained  by the spin-down  of an ultra-magnetized millisecond NS \citet{Stratta:2018xza,Gompertz:2013aka,Wang:2024fun}. It has been demonstrated that the spin-down luminosity from multipolar EM emission can effectively account for the plateau phase in the majority of GRB afterglows offering the complex magnetic field structure of magnetars as one of the candidates for the GRB's central engine \citep{Wang:2024fun}.

X-ray flares are also frequent events occurring in about 33$\%$ of GRBs within a time interval ranging from 30 to $10^5$ s after the GRB trigger time \citep{Chincarini:2010bp}.  The appearance of early X-ray flares during the light curves has been widely considered previously \citep{Ruffini:2017xsr,Zheng:2021nex,Saji:2023hsi,Zhang_2006}. It has been suggested that the prolonged activity of the central engine in a GRB  is responsible for flaring features alongside afterglow emissions  \citep{burrows2005bright,fan2005late,Beniamini:2015fta,Wada_2024}. However, explaining the late-time flaring activities is challenging in terms of the BH central engine and are more in favor of the idea of rapidly rotating magnetar central engines for GRBs  \cite{Rowlinson:2010ux,Metzger:2007cd,Zhang_2001,Dai1998}. It has been investigated that the magnetohydrodynamic instabilities in a rotating NS can produce a strong temporary magnetic field of order $10^{17} G$, and the magnetic reconnection effects  lead to multiple X-ray flares \citet{Kluźniak_1998,DAI}.  Beyond gamma-ray and X-ray observations, there are various data from optical to radio wavelengths   that support the  magnetar central engine model \citet{Rowlinson:2013ue,Xie_2020,Lazarus:2011cx}. These observations offer high values of the magnetic field of the order of $B\sim10^{12}G-10^{17}G$ with multipolar  structures \cite{Wang:2024fun}.  High  values of the quadrupole magnetic field component up to $\sim10^3$ times greater than dipole magnetic field have been proposed to explain GRB features \citet{Ruffini_2018,Stratta:2018xza}.

The multimessenger nature of magnetars through GW observations provides constraints on the deformation parameter, ellipticity, to be of order $\sim 10^{-5}-10^{-2}$ \rm   \cite{Lasky:2015olc}. The GW signal generated by magnetars can be explored more precisely by current and upcoming GW detectors \citet{Lai:1994ke,Sur:2020imd,Xie:2022igk}. These signals are generated primarily due to the extreme magnetic fields and rotational instabilities that cause deviations from axisymmetry.  Such observations will provide valuable insights into the physics of dense matter and the strong gravitational fields characteristic of these highly magnetized NSs. In the case of short GRBs, it has been shown that X-ray afterglow observations can be applied to constrain the properties of post-merger remnants, providing essential insights into the nuclear EoS and the GW emissions generated by newly formed NSs \citep{SARIN1}. Moreover, the polarization characteristics of the EM counterpart of the GW created by coalescence of the binary sources in short GRBs have been considered previously in \citet{Shakeri:2018qal}.

In this paper, we consider the evolution of a deformed magnetar as a power source for GRBs. We use the Maclaurin spheroid to model a newborn magnetar to explain different properties of  both EM and GW observations  of GRBs. Maclaurin spheroid as a homogeneous rotating object was first introduced by 
\citet{S.Chandrasekhar:1969} and later on used to model the spin-down of pulsars by \citet{SHAPIRO1990}, where the the evolution of eccentricity leads to spin variation of NS. Despite previous studies \citet{Xie:2022igk,Lasky:2015olc,Suvorov:2020eji,Rueda:2022mwh} in which the time evolution of angular velocity is derived directly from rotational energy loss of magnetar, here, we examine angular momentum as a function of the Maclaurin spheroid's eccentricity, which evolves over time. Our approach is more fundamental since rotational evolution originates from the deformation of the new NS, which can result in both spin down and spin up of the object. It is demonstrated that in the absence of accretion, during the loss of angular momentum, new born NS can also experience spin-up \citet{SHAPIRO1990}, it has been argued that this process eventually leads to the formation of a BH.   However, in this paper, we examine the stability regions within the parameter space more accurately, where the spin-up is caused by the decreasing moment of inertia due to the softness of the EoS, and  BH formation does not occur.

The spin-up mechanism we discuss differs fundamentally from the FRED-shaped light curves observed in many long GRBs, which are generally interpreted as evidence of the central engine spinning down \citep[see, e.g.,][]{2012PThPh.127..331P}. In our model, the spin-up typically manifests during the early afterglow phase, appearing at the end of the prompt emission  and is associated with the early X-ray flare activity. This is in contrast to the FRED profile, which characterizes the early prompt phase and reflects spin-down behavior.

We investigate the impact of key model parameters, such as the magnetic field, quadrupole-to-dipole ratio, mass and EoS, on both EM and GW emissions. We assume that both dipole and quadrupole emissions contribute to the EM radiation of the magnetar. We evaluate different features of the X-ray light curves under the influence of the properties of the system, such as the total mass of the progenitor, the magnetic field of the newborn magnetars, for both compressible and incompressible  fluids. Moreover, in the case of compressible fluids, we consider the impact of various polytropic (adiabatic) indices on the luminosity features. We see that for some specific values of the polytropic index, a peak in the luminosity appears, which will fade with time, can provide an explanation for the X-ray flares in GRBs.

The present paper is organized as follows. In Sec. \ref{se2}, we review the main features and properties of the Maclaurin spheroid in the context of both incompressible and compressible fluids. Then, we consider the evolutionary equation for the eccentricity due to the variation in angular momentum for both types of fluids mentioned. In Sec. \ref{se3}, we discuss the evolution of the NS's dipole, quadrupole, and total EM luminosity due to the evolution of eccentricity caused by angular momentum loss.  In Sec. \ref{se4},  we evaluate the GW radiation of a new born NS and examine the detectability of the resulting GW emission in the light of future  detectors.
In Sec. \ref{se5}, we present the total energy loss of the magnetar during spin-down due to the evolution of the eccentricity, and finally,
in Sec. \ref{sec6}, we discuss our results and  give some concluding remarks. In this paper, we are working in geometrical units in which $G=c=\hbar=1$.

\section{FRAMEWORK}\label{se2}
\begin{figure}

	\includegraphics[width=\columnwidth]{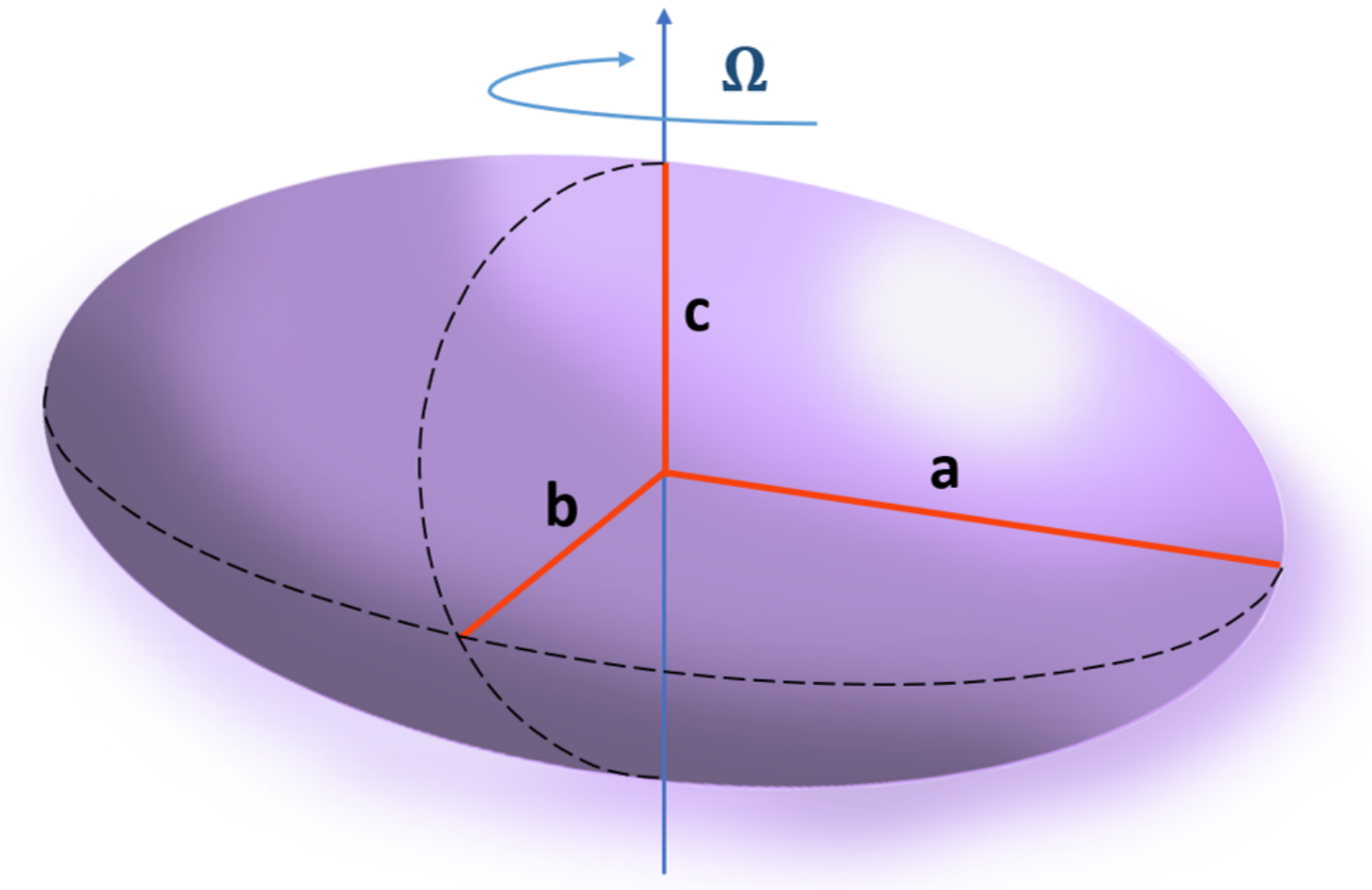}
    \caption{A schematic view of an asymmetric NS that rotates around $\hat z\rm-{axis}$,  the ellipticity $\epsilon$ and eccentricity e are described by Eqs (\ref{eq:3}) and (\ref{eq:4}), respectively. }
    \label{fig1}
\end{figure}

In this section, we present the Maclaurin spheroid as a model for the central engine of GRBs.
The angular velocity of Maclaurin spheroid is a function of  eccentricity, which measures the deformation of NS in the polar plane and is defined as \citet{S.Chandrasekhar:1969,SHAPIRO1990} :
\begin{align}\label{eq:3}
    e=(1-\frac{c^2}{a^2})^{1/2} ,
\end{align}
where a and c are shown in Fig. \ref{fig1} with c coinciding the rotational axis of the NS. 
In order to consider GW emissions, we assume an triaxial configuration of the NS with small deformation in the equatorial plane so that ellipticity is defined as
\begin{align}\label{eq:4}
    \epsilon&=\frac{a-b}{(a+b)/2},
\end{align}
with equatorial semi-axes a and b shown in Fig. \ref{fig1}. For a spheroid with density $\rho$, the angular velocity is given by
\begin{align}\label{eq:1}
    \Omega=\frac{2\pi}{P}=(2\pi\rho g(e))^{1/2},
\end{align}
with 
\begin{align}
    g(e)=\frac{(1-e^2)^\frac{1}{2}}{e^3}(3-2e^2)\arcsin(e)-\frac{3(1-e^2)}{e^2}.
\end{align}

The Maclaurin spheroid has angular momentum and mass defined as
\begin{align}\label{eq:5}
    J=I \Omega=\frac{2}{5}M a^2 \Omega,
\end{align}
\begin{align}
    M=\frac{4\pi}{3}\rho a b c\approx\frac{4\pi}{3}\rho a^3 (1-e^2)^{1/2},
\end{align}
where I is the moment of inertia. Here, due to the smallness of $ \epsilon$,  we take $a\approx b$. 
It has been shown that for a  equilibrium configuration with constant mass, entropy, and chemical potential,  the variation in energy, E, is related to the variation of angular momentum, J, according to \citet{1969ApJ...157.1395O,SHAPIRO1990} :
\begin{align}\label{eq:12}
    dE=\Omega dJ.
\end{align}
This relation can interpret the transformation of angular momentum into EM and gravitational energy emissions. During the spin-down, eccentricity evolves from a maximum value to a small value, making the star less oblate. Maximum eccentricity for a Maclaurin spheroid, known as the bifurcation point of a Maclaurin and Jacobian sequence, is approximately 0.813 \citet{S.Chandrasekhar:1969}, and at this point, bar-mode instabilities emerge. To avoid these dynamical instabilities, we consider the maximum eccentricity value at the bifurcation point.
In fact, oscillations of a perturbed NS have different modes among which f-modes and r-modes are the most susceptible to the Chandrasekhar-Friedman-Schutz (CFS) instability, leading to GW emissions \cite{PhysRevD.22.249,Chandrasekhar:1970pjp,Chugunov:2018isi}.
The instabilities induced by f-modes can be excited for relatively high rotational speed $\Omega\gtrsim0.8\ \Omega_{k}$ \cite{Friedman:1989zzb}, where  $\Omega_{k}$ is the angular frequency corresponding to the mass-shedding limit, given by 
\begin{align}
    \Omega_k\approx C\sqrt{\left(\frac{M}{M_{\odot}}\right)\left(\frac{10 km}{a}\right)^3},
\end{align}
where $C= 7800 s^{-1}$. This approximate frequency is obtained for a set of EoSs, marking the point beyond which the centrifugal force would tear apart the NS \cite{Haensel:1989mvc,Cook:1993qj,Kruger:2021zta,Andersson:2002ch}. On the other hand, r-mode toroidal oscillations, which are triggered by the Coriolis force, are unstable due to GW radiation even for slowly rotating perfect fluid stars \cite{Andersson:1997xt}.  While r-modes offer a larger instability parameter space, this window will be narrowed down due to the bulk/shear viscosity at high/low temperatures giving rise to a spin threshold of around the sub-kHz regime \cite{Glampedakis:2017nqy,Kraav:2024cus}. The detectability of the  r-mode instability is discussed in \cite{Kokkotas:2015gea,Caride:2019hcv,Mytidis_2015}. It was argued that the estimated rates for these events are 
not high enough to make detection very likely with second-generation GW detectors.

In the following, we consider the Maclaurin spheroid with two different types of fluids, (i) incompressible and (ii) compressible,  to model a newborn magnetars as the central engine of GRBs. We focus on the  secularly stable spheroids with $0\leq  T/|W|\leq 0.13$, where $T$ and $W$ are the rotational energy and  gravitational binding energy,  respectively. This condition restricts  the rotational frequency to be well below the mass-shedding limit; therefore, f-modes can not be excited in this regime. In the case of r-mode instabilities, in spite of their potential, we will not consider them in this paper.

\subsection{Incompressible Model}
Here, we introduce the incompressible Maclaurin spheroid and it's evolutionary equations. In this model, the density of the fluid forming the spheroid is constant; therefore, the semimajor axis, $a$, and moment of inertia, $I$, can be written as a function of eccentricity,
\begin{align}\label{eq:7}
    a&=a_{0}(1-e^2)^{-1/6},     & a_{0}&=(\frac{3M}{4\pi\rho})^{1/3},\nonumber\\
    I&=I_{0}(1-e^2)^{-1/3},      & I_{0}&=\frac{2}{5}M a_{0}^2,
\end{align}
where $a_{0}$ and $I_{0}$ are the radius and inertial moment of a spherical object, respectively, assuming constant homogeneous density. According to these equations, the time evolution of the eccentricity induces a temporal changes in both $a$ and $I$. 
Assuming that mass remains unchanged during the evolution, one can find the following  equation governing the variation of total angular momentum  $J$ by taking the logarithmic derivative of Eqs. (\ref{eq:1}), (\ref{eq:5}) and (\ref{eq:7}) :

\begin{align}\label{eq:11}
   \frac{\dot{J}}{J} = \frac{1}{2} \left( \frac{g'}{g} + \frac{4}{3} \frac{e}{1-e^2} \right) \dot{e},
\end{align}

where $g^{'}\equiv dg/de$. 
In next subsection, we discuss a more general form of the Maclaurin spheroid with a polytropic density profile.

\subsection{Compressible model}
In order to consider compressible  Maclaurin spheroids, we take polytropic EoS as

\begin{align}
    P=K\rho^{\Gamma},
\end{align}
where $P$ is pressure and  $\Gamma=1+1/n$ and $n$ is the polytropic index. By solving Poisson's equation for a uniformly rotating spheroid in hydrostatic equilibrium   and applying proper symmetry and boundary conditions, one can get  \citet{shapiro1983black} 
\begin{align}\label{eq:16}
    \left(\frac{\rho}{\rho_{0}}\right)^{\Gamma-4/3}=\frac{3}{2}(1-e^2)^{2/3}A_{3}(e),
\end{align}
where $\rho_{0}$ is the constant density of a sphere ($e=0$) with radius $a_{0}$ and $A_{3}(e)$ is a function of the eccentricity, which is given by
\begin{align}
    A_{3}(e)=\frac{2}{e^2}\left[1-(1-e^2)^{1/2}\frac{\arcsin{e}}{e}\right].
\end{align}

Note that in the limit of large $\Gamma$, the compressible model of the fluid tends to the incompressible model. Here, we a have generalized form of the set of Eq. (\ref{eq:7}) for the  compressible model:
\begin{align}\label{eqq13}
     a&=a_{0}\left(\frac{\rho_{0}}{\rho}\right)^{1/3}(1-e^2)^{-1/6},    \nonumber\\
    I&=I_{0}\left(\frac{\rho_{0}}{\rho}\right)^{2/3}(1-e^2)^{-1/3},     
\end{align}
Following the same strategy used for the incompressible model and taking the logarithmic derivative of Eq. (\ref{eq:16}) one can have:
\begin{align}\label{eq:18}
    (\Gamma - 2) \frac{d\rho}{\rho} = 2 \frac{da}{a} + \left( \frac{A'_{3}}{A_{3}} - \frac{2e}{1 - e^2} \right) de.
\end{align}

where $A_{3}^{'}\equiv dA_{3}/de$. By substituting  Eq. (\ref{eq:18}) into the logarithmic derivation of Eqs (\ref{eq:5}) regarding $\dot M=0$, the following equation is obtained
\begin{align}\label{eq:19}
    \frac{\dot{J}}{J} = \frac{1}{2} \bigg\{ \frac{g'}{g} + \frac{1}{3\Gamma - 4} \left[ \frac{4(\Gamma - 1)e}{1 - e^2} - \frac{A'_{3}}{A_{3}} \right] \bigg\} \dot{e}.
\end{align}
\begin{align}\label{eq:19}
   \eta=ae/r
\end{align}

 Knowing the EM and GW emissions, through Eqs.(\ref{eq:5}) and (\ref{eq:12}) one can eliminate $J$ and $\dot{J}$ from Eqs. (\ref{eq:11}) and (\ref{eq:19}) and find the evolutionary equation for eccentricity for both the compressible and incompressible models.
\section{EM Radiation and Spin-down}\label{se3}

In this section, we consider EM radiation from a magnetar to determine the evolution of eccentricity and then find the EM light curve. In subsection \ref{ss1}, we focus on dipole radiation, followed by calculations for quadrupole radiation in subsection \ref{ss2}, and the (dipole + quadrupole) total EM radiation in subsection \ref{ss3}.

\subsection{Dipole luminosity}\label{ss1}
Energy loss via EM dipole moment (dipole luminosity) is given by
\begin{align}\label{eqq16}
       L_{\rm dip}&= -\dot{E}_{\rm dip}=\frac{2}{3}B_{ \rm dip}^2a_{0}^6{\sin^2}{\alpha} \Omega^4\\ \nonumber
    & =  8.13 \times 10^{49} \rm erg/s 
\left(\frac{a_0}{10^6 cm}\right)^6\left(\frac{0.5 ms}{P}\right)^4 \left(\frac{B}{10^{15}G }\right)^{2},
\end{align}

where $P=2\pi/\Omega$, $B_{\rm dip}$  is the dipole magnetic field and $\alpha$ is the angle between the magnetic dipole moment and the rotational axis. Hereafter, we select $\alpha=\pi/2$. In Eq. (\ref{eqq16}) we use appropriate values for scaling parameters. Combining  Eqs (\ref{eqq16}) and (\ref{eq:12}) the following equation for the angular momentum evolution is evaluated:
\begin{align}\label{eq:21}
    \dot{J}_{dip}=-\frac{2}{3}B_{dip}^2a_{0}^6\Omega^3=-\beta\Omega^3.
\end{align}
Finally combining Eqs (\ref{eq:5}), (\ref{eq:11}) and (\ref{eq:21}), evolutionary equation for eccentricity in incompressible fluid is
\begin{align}\label{e18}
    \frac{de}{dt} = -\frac{\Omega^4(e) \beta}{\pi I_0 \rho} (1 - e^2)^{1/3} \left[ g'(e) + \frac{4}{3} \frac{g(e) e}{1 - e^2} \right]^{-1}.
\end{align}

On the other hand, by using Eqs  (\ref{eq:5}), (\ref{eq:19})  and (\ref{eq:21}) for the compressible fluid can be obtain as
\begin{align}\label{e19}
    \frac{de}{dt} &= -\frac{\Omega^4(e) \beta}{\pi I_0} \left( \frac{\rho_{0}}{\rho^4} \right)^{1/3} (1 - e^2)^{1/3} \times \\
    &\bigg\{ g'(e) + \frac{g(e)}{3\Gamma - 4} \left[ \frac{4(\Gamma - 1)e}{1 - e^2} - \frac{A'_{3}}{A_{3}} \right] \bigg\}^{-1}.
\end{align}

\subsection{Quadrupole Luminosity}\label{ss2}
Energy loss for EM quadrupole magnetic moment (quadrupole luminosity) is given by
\begin{align}
    \dot{E}_{\rm quad}=-\frac{32}{135}B_{\rm quad}^2a_{0}^8{\sin^2}{\beta_{1}}({\cos^2}{\beta_{2}}+10{\sin^2}{\beta_{2}})\Omega^6,
\end{align}
where $B_{\rm quad}$ is the quadrupole magnetic field and  $\beta_{1}$ and $\beta_{2}$  are the inclination angles of the magnetic moments. It is instructive to introduce $\kappa$
as  the quadrupole-to-dipole ratio:
\begin{align}
    \kappa = \sin^2 \beta_{1} \left( \cos^2 \beta_{2} + 10 \sin^2 \beta_{2} \right) \frac{B_{\rm quad}^2}{B_{\rm dip}^2}.
\end{align}

An order of magnitude estimation for EM quadrupole luminosity is given by
\begin{align}
     &L_{\rm quad}=\frac{32}{135}B_{\rm dip}^2a_{0}^8\kappa\Omega^6\\ \nonumber
    &=5.08 \times 10^{51} erg/s \left(\frac{B}{10^{15} G}\right)^2\left(\frac{a_0}{10^6 cm}\right)^8\left(\frac{0.5 \rm ms}{P}\right)^6 \left(\frac{\kappa}{500}\right).
\end{align}
Therefore, angular momentum loss due to quadrupole magnetic moment radiations is
\begin{align}
    \dot{J}_{\rm quad}=-\xi\Omega^5,
\end{align}
where $\xi=-32/135B_{\rm quad}^2a_{0}^8\kappa$, leading to the evolutionary equation for the eccentricity of the incompressible model as
\begin{align}
    \frac{de}{dt} = -\frac{\Omega^6(e) \xi}{\pi I_0 \rho} (1 - e^2)^{1/3} \left[ g' + \frac{4}{3} \frac{eg}{1 - e^2} \right]^{-1}.
\end{align}

For  the compressible model, this evolutionary equation is evaluated as
\begin{align}
    \frac{de}{dt} &= -\frac{\Omega^6(e) \xi}{\pi I_0} (1 - e^2)^{1/3} \left( \frac{\rho_{0}}{\rho^4} \right)^{1/3} \times \\ \nonumber
    &\bigg\{ g' + \frac{g}{3\Gamma - 4} \left[ \frac{4(\Gamma - 1)e}{1 - e^2} - \frac{A'_{3}}{A_{3}} \right] \bigg\}^{-1}.
\end{align}

\subsection{Total EM Luminosity}\label{ss3}
Now, we consider the total EM radiation in order to determine the angular momentum loss, which is expressed as
\begin{align}
    \dot{J}_{\rm EM}=\dot{J}_{\rm dip}+\dot{J}_{\rm quad}=-\beta\Omega^3-\xi\Omega^5.
\end{align}
Therefore, the following equations describe the evolution of eccentricity for the incompressible fluid as 
\begin{align}
    \frac{de}{dt} = -\frac{\Omega^4(e) \beta}{\pi I_0 \rho} (1 - e^2)^{1/3} \left( 1 + \frac{\Omega^2(e) \xi}{\beta \rho} \right) \left[ g' + \frac{4}{3} \frac{eg}{1 - e^2} \right]^{-1},
\end{align}

and in the case of the compressible model, we obtain 
\begin{align}
     \frac{de}{dt} &= \frac{-\Omega^4(e)\beta}{\pi I_0} (1-e^2)^{1/3} \left( \frac{\rho_{0}}{\rho^4} \right)^{1/3} \left( 1 + \frac{\Omega^2(e)\xi}{\beta \rho} \right) \times \\ \nonumber
    & \bigg\{ g' + \frac{g}{3\Gamma-4} \left[ \frac{4(\Gamma-1)e}{1-e^2} - \frac{A'_{3}}{A_{3}} \right] \bigg\}^{-1}.
\end{align}

\begin{figure}
\includegraphics[width=\columnwidth]{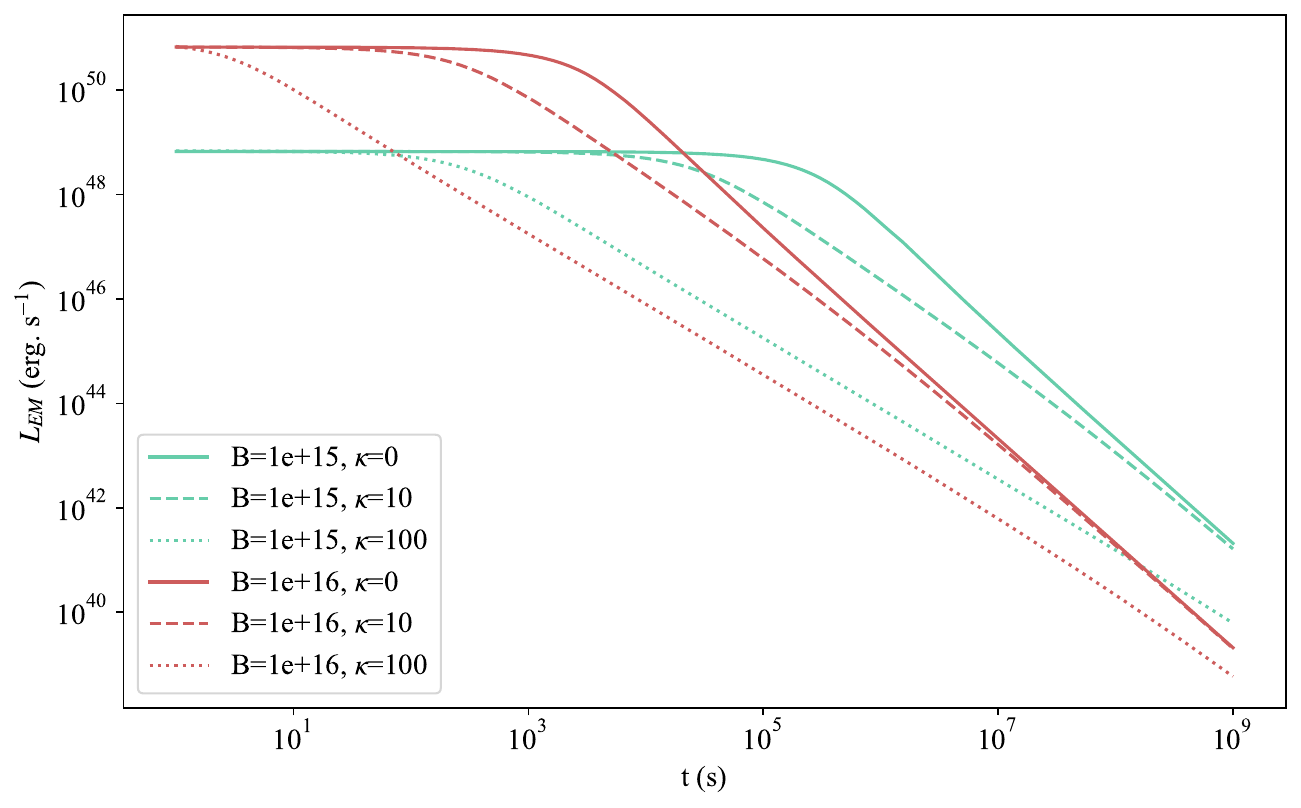}
    \caption{ Total EM luminosity for the incompressible model with fixed mass $M_{T}=2M\odot$. Different magnetic fields and quadrupole-to-dipole ratios are labeled.}
    \label{fi2}
\end{figure}
\begin{figure}
\includegraphics[width=1.1\columnwidth]{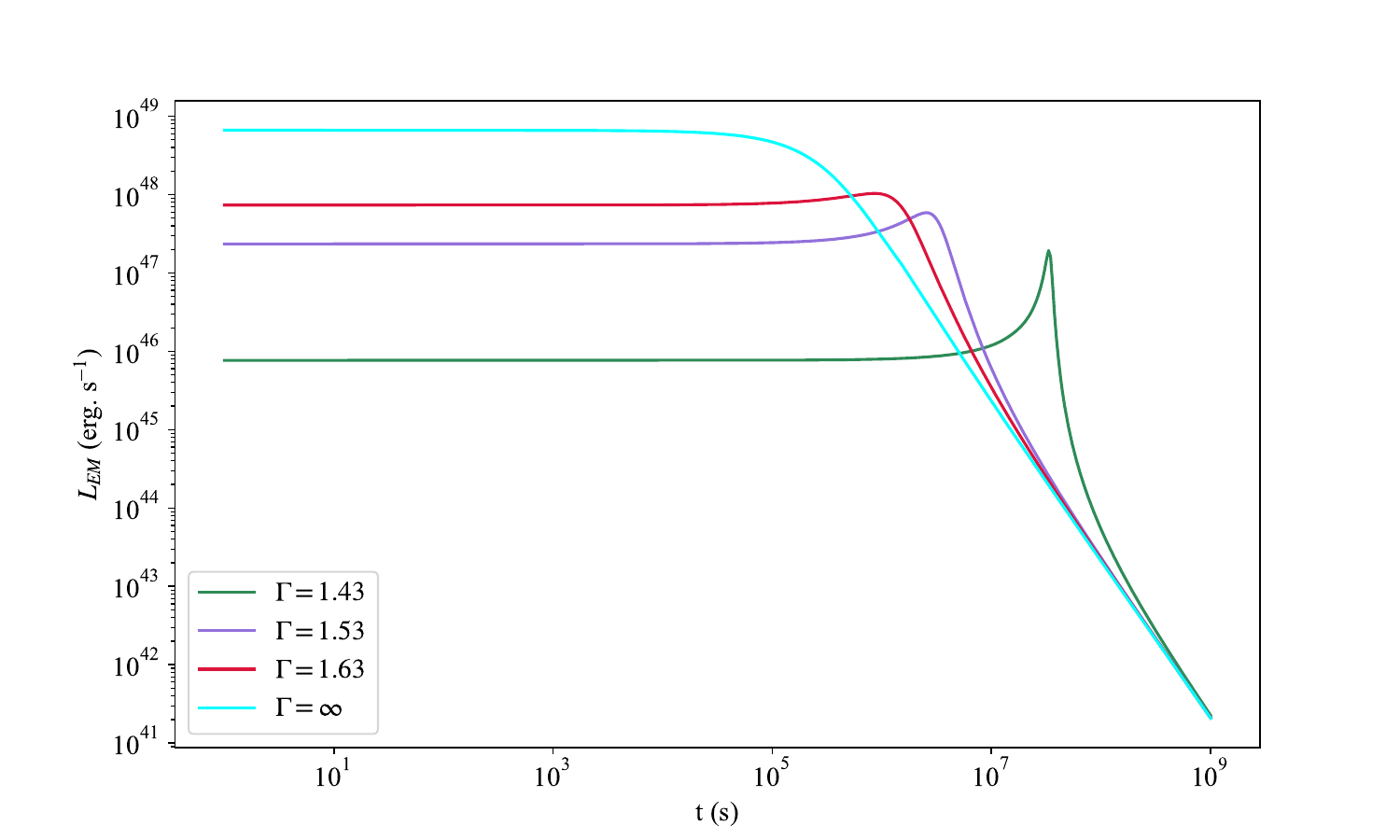}
    \caption{EM dipole luminosity for the incompressible and compressible  models. Here, we assume that $M_{T}=2M\odot$ and $B=10^{15}G$. In the case of the compressible model, different adiabatic indexes are labeled.}
    \label{fi3}
\end{figure}
\begin{figure}

	\includegraphics[width=1.1\columnwidth]{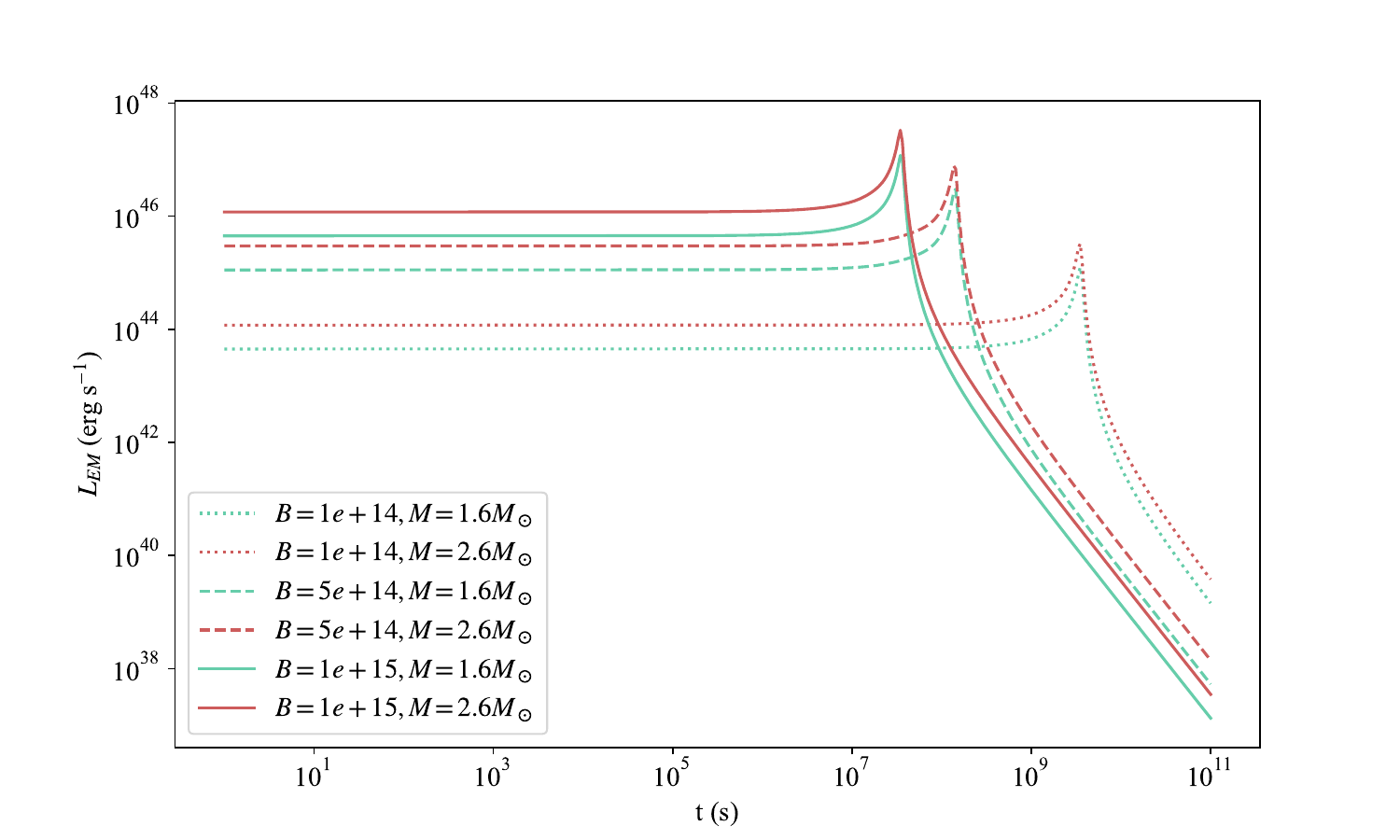}
    \caption{EM dipole luminosity for the compressible model is plotted for $\Gamma=1.43$, where the magnetic field  and mass of the magnetars are labeled.}
    \label{fi4}
\end{figure}

\begin{figure}

 \includegraphics[width=1.1\columnwidth]{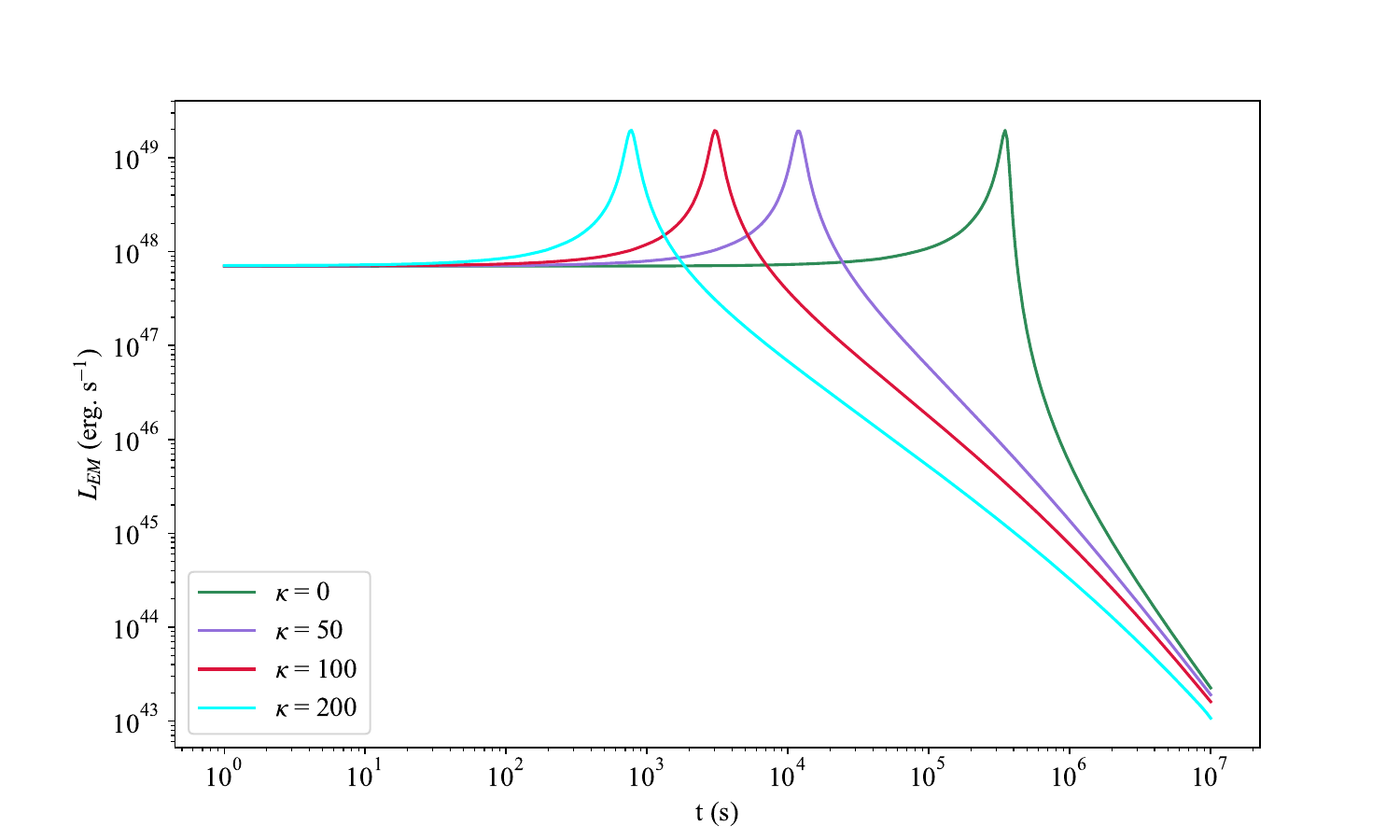}
    \caption{EM dipole and quadrupole emission of the compressible model 
with  $\Gamma=1.43$, $B=10^{16} G$ and different fractions of dipole to quadrupole moments as labeled.}
    \label{fi5}
\end{figure}
Fig. \ref{fi2} shows the total EM luminosity for a magnetar containing incompressible fluid. Here, we assume that the dipole magnetic field strength is $B=10^{14} - 10^{16} \rm G$ and the quadrupole is up to $500$ times larger than the dipole moment. It is seen that for higher ratios of magnetic  quadrupole-to-dipole moments $\kappa$, the EM light curve of the  magnetar starts to decay at earlier times with a lower slope. This is a result of spin-down with a faster rate at earlier times. 
We find that the luminosity is mainly powered by the dipole magnetic field at the earlier times, and at later times, it is dominated by dipole emission.


In Fig. \ref{fi3}, the effect of different polytropic EoSs on the  dipole EM luminosity is investigated. It is illustrated that for some specific values of adiabatic index $\Gamma$, close enough to 4/3, a peak in the light curve appears. As can be seen, the peak fades away as $\Gamma$ increases and goes beyond $4/3$. These  regions within the parameter space where the spin-up is caused by the decreasing moment of inertia due to the softness of the EoS, and  BH formation does not occur,  we attribute the resulting flaring behavior in the light curve to these regions. The flaring behavior is closely tied to the rotational dynamics of new born NS, which leads to  a temporary spin-up of the NS during its early spin-down phase. This effect  crucially depends on the applied EoS and the adiabatic index $\Gamma$, as can be seen from Eq. (\ref{e19}). The non-axisymmetric deformations of the NS, including a compressible fluid for $\Gamma\approx 4/3$ will show a nontrivial behavior leading to a temporary spin-up behavior that is reflected in the  observed  flares upon the light curves. 

The incompressible model is parameterized with  $\Gamma=\infty$ and as it is seen in Fig.  \ref{fi3}, it has the highest maximum initial luminosity for  a given magnetic field and mass of the magnetar, 
and the maximum initial luminosity increases as softer EoSs are applied.
Additionally, one can see that for stiffer EoSs, the decay time in the light curve occurs at earlier times. Note that the late-time behavior of the light curve for all models with different $\Gamma$ values is similar. In Fig. \ref{fi4}, we explore the effect of both mass and magnetic field on the light curve of the dipole EM radiation for a compressible fluid with $\Gamma=1.43$. As shown in Fig. \ref{fi3}, the magnetic field has a significant impact on the peak luminosity, $L_{\rm p}$ and the peak time, $T_{\rm p}$, while the mass's effect  in a given range on $L_{\rm p}$ and $T_{\rm p}$ is not significant. The allowed range of the parameter space of mass and $\Gamma$ is determined through the stability analysis of the magnetar under deformation (see Appendix (\ref{ap1})).  We introduce the valid mass range for our model based on black hole formation criteria following the method presented in \citet{SHAPIRO1990,wheeler1972magic}

In Fig. \ref{fi5}, the magnetic dipole luminosity is compared to the total EM luminosity, including quadrupole emission with  different $\kappa$ in the compressible model. We find that taking the quadrupole magnetic moment into account leads to an earlier peak time in the light curves.  Moreover, for a given dipole magnetic field, the effect of the quadrupole-to-dipole moment ratio on the peak luminosity is negligible.

\section{GW Radiation and Spin-down}\label{se4}
\label{sec:maths} 

\subsection{GW Luminosity}
In this section, we investigate the GW radiation from a deformed magnetar due to the asymmetry in its equatorial plane. The GW luminosity of a deformed magnetar is expressed as \citet{Andersson:2019yve}
\begin{align}\label{eq:37}
   &L_{\rm GW} = - \dot{E}_{\rm GW} = \frac{1}{5}\left<\frac{d^3{I}_{jk}}{dt^3}\frac{d^3{I}_{jk}}{dt^3}\right> 
   = \frac{32}{5} I^2 \epsilon^2 \Omega^6 \nonumber \\ 
   &= 1.73 \times 10^{50} \, \text{erg/s} \left(\frac{a_0}{10^6 \, \text{cm}}\right)^4 \left(\frac{\epsilon}{10^{-3}}\right)^2 \left(\frac{0.5 \, \text{ms}}{P}\right)^6 \left(\frac{M}{2 M_{\odot}} \right)^{2}.
\end{align}

By introducing $\gamma_0\equiv32I_0^2\epsilon^2/5$, the time evolution of the eccentricity due to GW radiation from an incompressible fluid is given by the following equation 
\begin{align}
    \frac{de}{dt} = -\frac{\Omega^6(e) \gamma_0}{\pi I_0 \rho} (1 - e^2)^{-1/3} \left[ g'(e) + \frac{4}{3} \frac{g(e) e}{1 - e^2} \right]^{-1}.
\end{align}

and for the compressible fluid one has
\begin{align}
     \frac{de}{dt} &= -\frac{\Omega^6(e) \gamma_0}{\pi I_0 \rho} \left( \frac{\rho_{0}}{\rho} \right)^{5/3} (1 - e^2)^{-1/3} \times \\ \nonumber
     & \bigg\{ g'(e) + \frac{g(e)}{3\Gamma - 4} \left[ \frac{4(\Gamma - 1)e}{1 - e^2} - \frac{A'_{3}}{A_{3}} \right] \bigg\}^{-1}.
\end{align}

\begin{figure}

    \centering	\includegraphics[width=1.1\columnwidth]{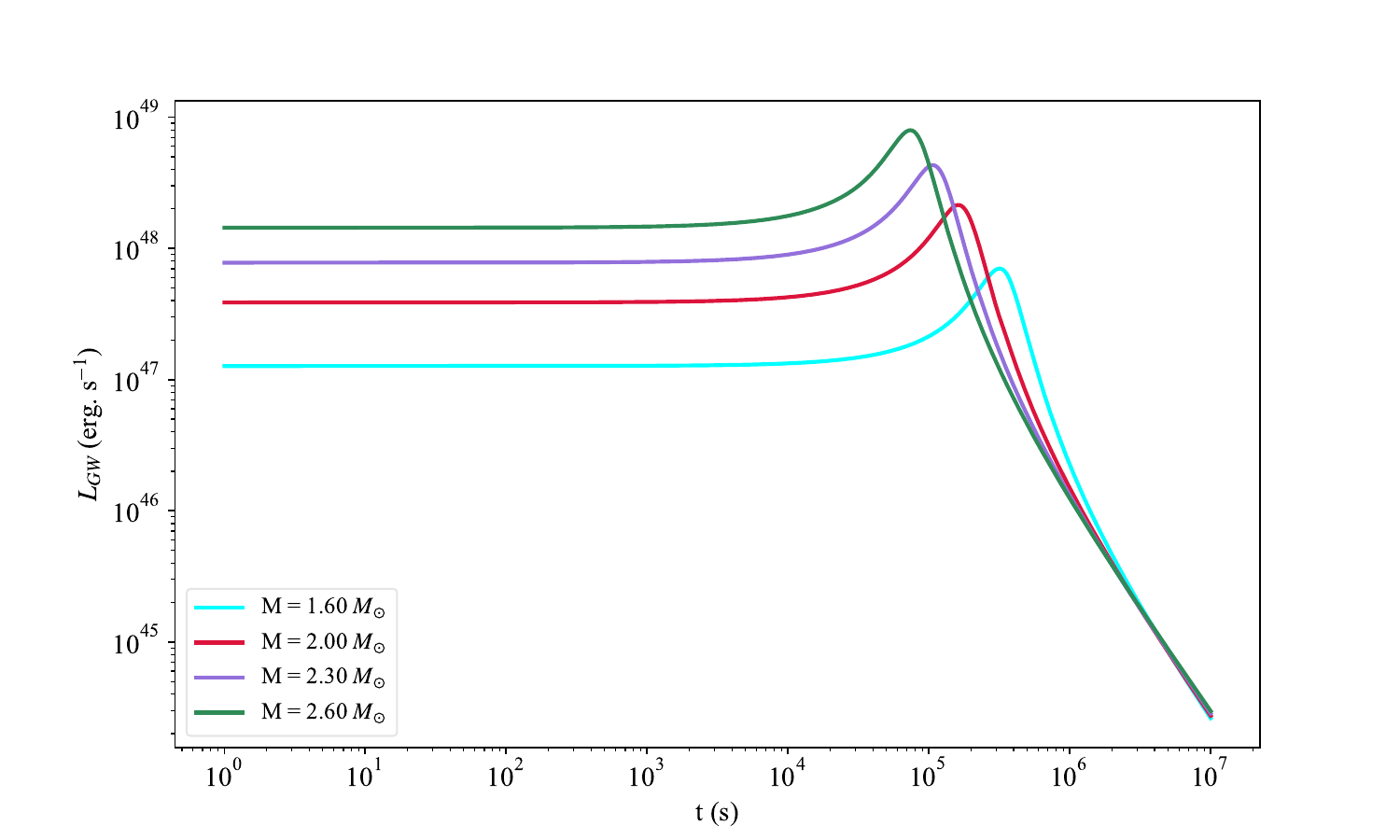}
    \caption{GW luminosity for the compressible model 
with  $\Gamma=1.43$, and $\epsilon=10^{-3}$ for different masses of NSs as labeled.}
    \label{f6}
\end{figure}

\begin{figure}

	\includegraphics[width=\columnwidth]{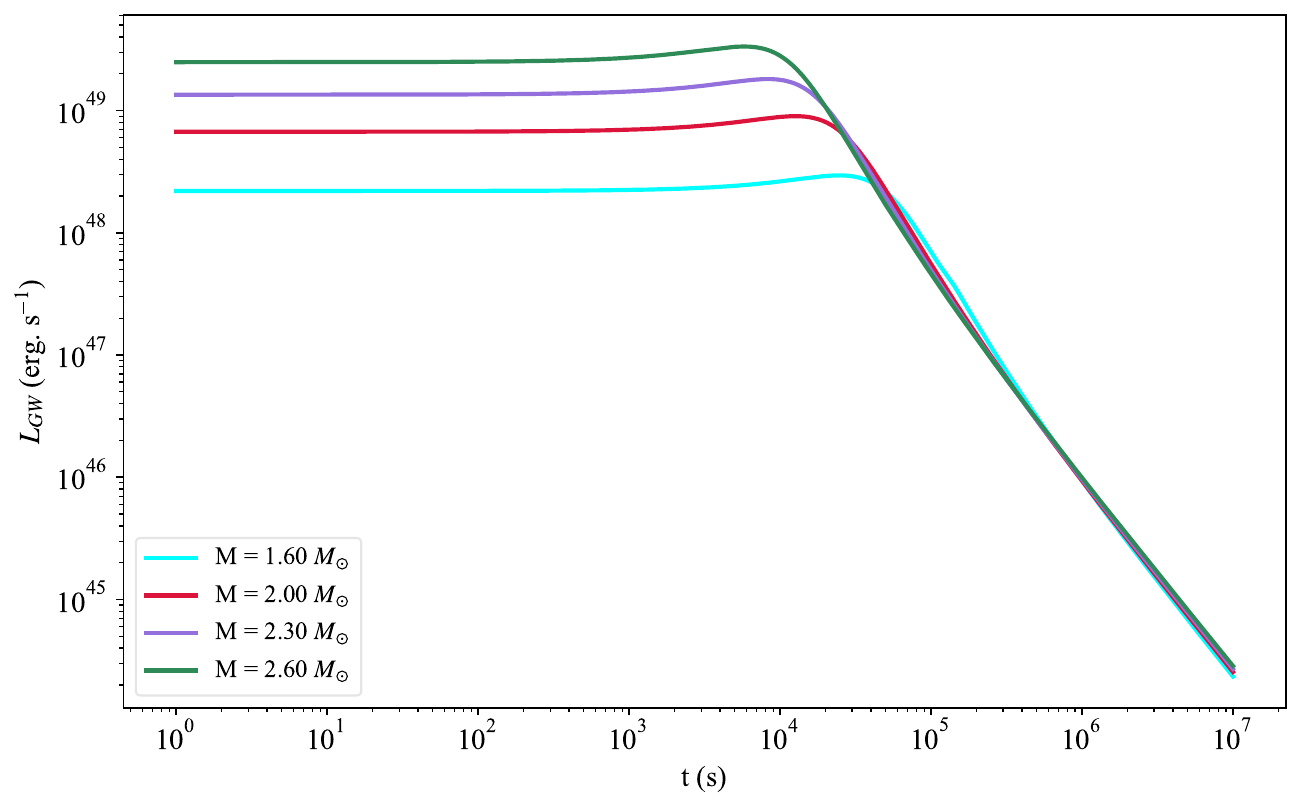}
    \caption{GW luminosity for the compressible model 
with  $\Gamma=1.53$, and $\epsilon=10^{-3}$ for different masses of NSs as labeled.}
    \label{f7}
\end{figure}

\begin{figure}
	\includegraphics[width=\columnwidth]{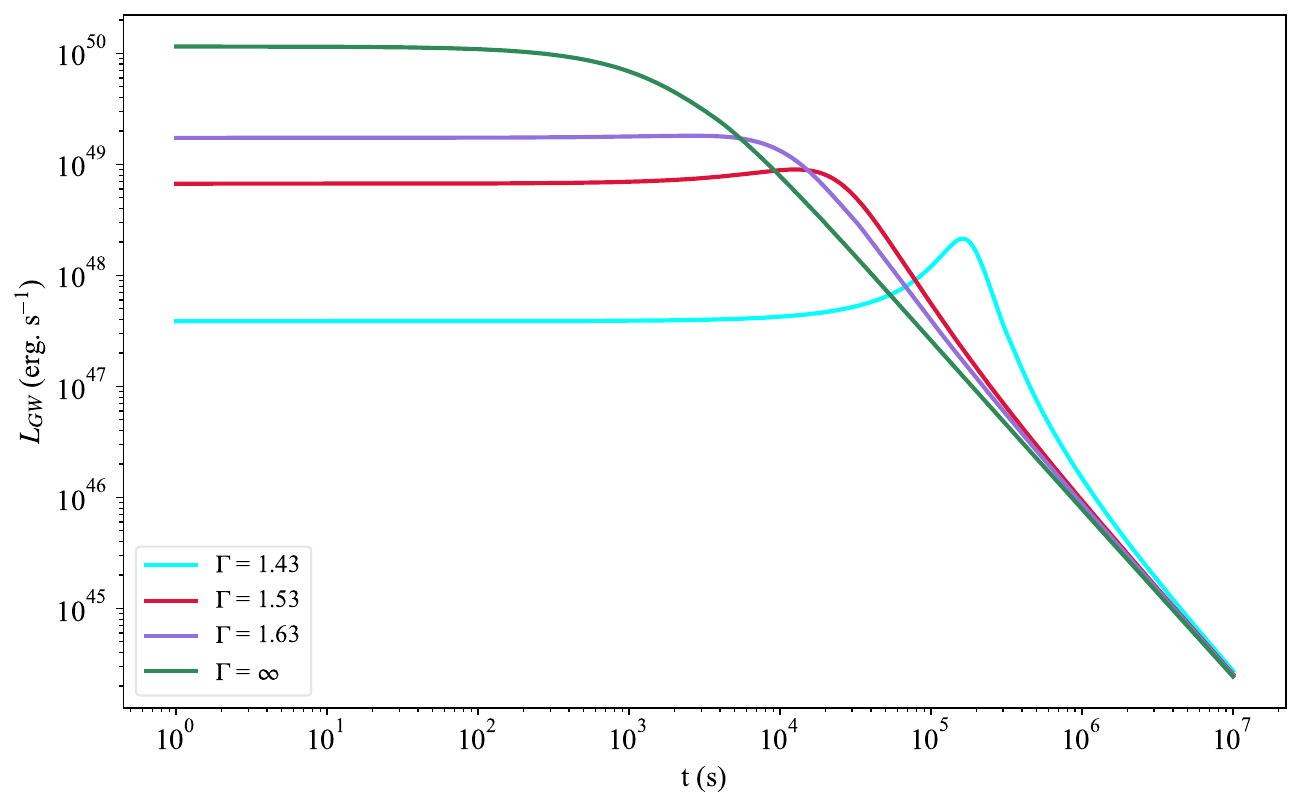}
    \caption{GW luminosity for the incompressible ($\Gamma=\infty$) and compressible  models. Here, we assume that $M_{T}=2M\odot$ and $\epsilon=10^{-3}$, in the case of the compressible model, different adiabatic indexes ($\Gamma$) are labeled.}
    \label{f8}
\end{figure}

We plot $L_{GW}$ for some cases in Fig. \ref{f6} - \ref{f8} to examine the gravitational energy loss from a newborn magnetar. Fig. \ref{f6} and Fig. \ref{f7}  describe GW luminosity for the compressible model for $\Gamma=1.43$ and $\Gamma=1.53$, respectively. These figures indicate that for higher masses, the luminosity peak appears at earlier times with a higher magnitude. At late times, luminosity curves for all masses coincide. It is found that a higher amplitude signal might be detected for a magnetar with softer EoS during peak time. Fig. \ref{f8}  illustrates the GW luminosity for different EoSs. As $\Gamma$ tends to large values far beyond $4/3\approx1.33$, the peak in $L_{GW}$ fades away,  which is the same result as seen in the EM light curve. Furthermore, the behavior of maximum luminosity with respect to changes in $\Gamma$ is similar.

\subsection{Detectability of the GW Signals}

\begin{figure}

	\includegraphics[width=1\columnwidth]{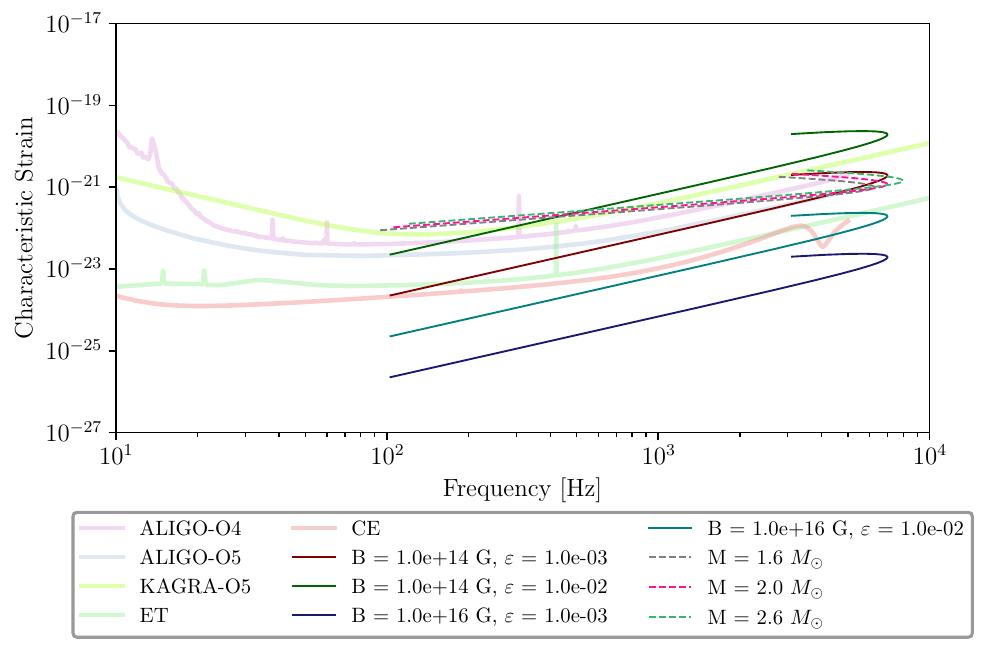}
    \caption{Sensitivities of GW detectors to GW characteristic strain produced by rotating magnetars for different masses, as labeled.}
    \label{fig9}
\end{figure}

Here, we consider the detectability of the GW signal originating from a deformed magnetar in the light of future GW detectors. The GW strain for the deformed magnetar located at distance D from the observer is given by \citet{Howell:2010hm,Bambi:2020tsh,Corsi:2009jt}
\begin{align}
    h(t)=&\frac{4}{D}{\epsilon}I\Omega(t)^2\ \approx5.31 \times 10^{-25}\times \\ \nonumber
  &\left(\frac{40 Mpc}{D}\right)\left(\frac{a_0}{10^6 cm}\right)^2\left(\frac{\epsilon}{10^{-3} }\right)\left(\frac{0.5 ms}{p}\right)^2\left(\frac{M}{2.6M_{\odot}}\right).
\end{align}
The characteristic GW strain as a dimensionless quantity can be calculated using
\begin{align}
    h_c=fh(t)\sqrt{dt/df}.
\end{align}
Knowing that $f=\Omega/\pi$ one can find that
\begin{align}
    h_c=\frac{4\pi^2}{D}{\epsilon}f^3\sqrt{dt/df}.
\end{align}
In order to find the explicit form of $h_c$ one needs to find the term $dt/df$, which can \citet{} be find according to
\begin{align}\label{e35}
    \frac{dE_{GW}}{df}=\frac{dE_{GW}}{dt}\frac{dt}{df}=L_{GW}\frac{dt}{df}.
\end{align}
Considering the domination of GW radiation at early times,
\begin{align}\label{e43}
    \frac{dE}{df} \approx  \frac{dE_{GW}}{df}= \pi^2 If,
\end{align}
which leads to the following characteristic GW strain:
\begin{align}
    &h_{c}(t) = \frac{1}{D}\sqrt{\frac{5If}{2}}= \frac{1}{D}\sqrt{\frac{5I\Omega(t)}{2\pi}}\\ \nonumber
  &\approx 1.45 \times 10^{-21}\left(\frac{40 Mpc}{D}\right)\left(\frac{a_0}{10^6 cm}\right)\left(\frac{0.5 ms}{P}\right)^{1/2}\left(\frac{M}{2.6M_{\odot}}\right)^{1/2}.
\end{align}
In order to find $h(t)$ one needs to determine $\Omega(t)$ due to variation of eccentricity, we plot   $ h_{c}$ in Fig. \ref{fig9} for different masses as labeled assuming $D=40 \text{Mpc}$, $a_{0}=10^6 cm$  and $P=0.5$ ms for a new born magnetar. It is seen that the variation in the mass in the given range has no significant impact on $h_{c}$ and the resulting curves for the three different values of masses almost coincide  with each other.

On the other hand, for later times, EM radiation dominates the energy loss, by rewriting  Eqs. (\ref{e35}) and (\ref{e43}) for the EM case, $dt/df$ is given as bellow:
\begin{align}
    \frac{dt}{df} \approx \frac{\pi^2If}{L_{EM}}.
\end{align}
The characteristic GW strain for this regime is obtained as
\begin{align}
     h_{c}(t) &=\frac{1}{D}\left(1+\frac{16}{45}\pi^2 a^2\kappa \Omega(t)^2\right)^{-1/2}\sqrt{\frac{24I^3\epsilon^2\Omega(t)^3}{\pi B_{dip}^2a^6}},
\end{align}
Here, we can ignore the effect of quadruple magnetic field, while $\kappa a^2 \Omega^2<<1$, and write $h_c(t)$ only for the EM dipole radiation:
\begin{align}\label{e40}
     h_{c}(t) =&\frac{1}{D}\sqrt{\frac{24I^3\epsilon^2\Omega(t)^3}{\pi B_{dip}^2a^6}}\approx 9.54 \times 10^{-22}\\ \nonumber
     &\left(\frac{40 Mpc}{D}\right)\left(\frac{0.5 ms}{P}\right)^{3/2}\left(\frac{M}{2M_{\odot}}\right)^{3/2}\left(\frac{\epsilon}{10^{-3}}\right)\left(\frac{10^{16}G}{B_{\text{dip}}}\right).
\end{align}
The characteristic GW strain for the case in which the spin-down is driven by the EM dipole radiation is plotted in Fig. \ref{fig9}  for various $B$ and $\epsilon$ as labeled. We see from Eq. (\ref{e40}) and Fig. \ref{fig9}  that 
a large ellipticity would produce a stronger GW signal. Eq. (\ref{e40}) is inversely proportional to the $B_{\text{dip}}$, however, the magnetic field should be large enough to reproduce  the observed  EM luminosity, therefore magnetic field strength has a lower bound. On the other hand, it has been shown that the large ellipticity can be generated thanks to high values of the magnetic field, which means larger magnetic field components  \citet{10.1093/mnras/stac859,Sur:2020imd,Lasky:2015olc,Stella:2005yz}.

The projected sensitivities for the ALIGO O4, ALIGO O5, KAGRA O5, Einstein Telescope (ET), and Cosmic Explorer (CE) detectors are also presented in Fig. \ref{fig9}. It is found that GW signals from magnetars with a specific set of parameters reach the sensitivity threshold of these detectors. 
Here, we report a new signature of a new born magnetar in the GW strain, where there is a turning point at high frequencies due to the spin-up  and spin-down for a range of compressible EoSs (here, we select $\Gamma=1.43$). This feature may be used to discriminate the GW signal generated by NS from those produced by BH central engine \cite{vanPutten:2024ftm}. We will consider this case in more detail in a follow-up paper.

In order to consider the detectability of the signal, we also calculate the root sum squared strain for GW as below (\citet{Abbott_2017}):
\begin{align}
    h_{rss}=\sqrt{2\int_{f_{\text{min}}}^{f_{\text{max}}}\left(|\tilde{h}_+(f)|^2+|\tilde{h}_{\times}(f)|^2\right) \, df}.
\end{align}
where $\tilde{h}_+(f)$ and $\tilde{h}_{\times}(f)$ are the plus and cross polarizations of GW, respectively. For the case of a deformed magnetar, we use the relations described in \citet{SHAPIRO1990} for $\tilde{h}_+$ and $\tilde{h}_{\times}$
polarizations, assuming the same contribution of both polarizations. The frequency range is evaluated from Eq. (\ref{eq:1}), $f=2\Omega$, for the compressible model to be $f_{min}=103 Hz $ and $f_{max}=6994 Hz$, which corresponds to the eccentricity changing from 0.81 to 0.005. Therefore, we have
\begin{align}\label{e43}
h_{rss}=&1.88\times 10^{-23} Hz^{-1/2}\\ \nonumber
&\left( \frac{\epsilon}{10^{-3}}\right) \left( \frac{40\text{Mpc}}{\text{D}}\right) \left( \frac{\text{M}}{2 M_{_{\odot}}}\right)^{1/2} \left( \frac{R}{10\ \text{km}}\right).
\end{align}
The signal-to-noise ratio (SNR) for a narrowband source at the central frequency $f_{0}$ is given by 
\begin{align}
    \frac{S}{N}=\Theta\frac{h_{rss}}{\sqrt{S(f_0)}}
\end{align}
where $\Theta$ is the  angle factor and for an optimally oriented source can be estimated as $\Theta_{rms}=\sqrt{2/5}$ \cite{Sutton:2013ooa}. Considering $S(f_{0}\approx 1\text{ kHz})$,  $h_{rss}$ is calculated for a given set of scaling parameters as in Eq. (\ref{e43}), and the SNRs for the current and future detectors are presented in Table \ref{t2}. This result shows that GW detectors might be able to detect GWs from newborn magnetars and record the specific signatures of spin evolution and deformation of the star.
\begin{table}[h]
    \centering
    \caption{Sensitivity and SNR values for different detectors.}
    \label{t2}
    \begin{tabular}{@{} l l l @{}}
        \hline
       GW Detector & $S(f_{0} \approx 1\text{ kHz})$$/ \sqrt{Hz}$ & SNR \\
        \hline\hline
        ALIGO O4  & $5.75\times10^{-24}$  & 2.02  \\
        KAGRA O5  & $1.28\times10^{-23}$  & 0.93 \\
        ALIGO O5  & $2.53\times10^{-24}$  & 4.69 \\
      ET        & $5.76\times10^{-25}$  & 20.62 \\
        CE       & $3.27\times10^{-25}$  & 36.32 \\
        \hline
    \end{tabular}
\end{table}


Note that being the GW strain  within the sensitivity range of the GW detector (as it is seen from Fig. \ref{fig9}) does not necessarily guarantee that the GW will be detected, though it suggests that detection should be possible under ideal conditions. In practice, the detector’s ability to identify a GW signal depends on several factors, such as noise, duration of the signal, and the methods used to search for the signal. If the SNR is too low (e.g., the signal is weak or buried in background noise), it may not be detectable.  For continuous or long-duration signals (such as those expected from deformed NSs or magnetars), noise can accumulate over time, and the signal can be buried in background noise, making detection harder despite being within the sensitivity range. Imperfect search algorithms or unknown signal characteristics could also result in a missed detection. Therefore, being in the sensitivity range is a necessary but not sufficient condition for the detection. However, if the GW signal associated with the EM light curve of the magnetar can be detected in the future, it would be a strong evidence that the magnetar can act as the central engine of GRBs.

\section{GW and EM spin-down}\label{se5}
This section describes the entire energy loss of the magnetar during EM and GW emissions. Despite the fact that the total luminosity is not an observable quantity by a single instrument, it can illustrate the dynamical behavior of a new born magnetar  through multimessenger emissions. 
\subsection{Dipole EM and GW Radiation}
Here, we consider spin down due to both dipole and GW radiations. The resulting evolutionary equation for eccentricity in incompressible is given by
\begin{align}
    &\frac{de}{dt} = \\ \nonumber
    & -\frac{\Omega^4(e) \beta}{\pi I_0 \rho} (1 - e^2)^{1/3} \left[ 1 + \frac{\Omega^2(e) \gamma_{0}}{\beta (1 - e^2)^{2/3}} \right] \left[ g'(e) + \frac{4}{3} \frac{g(e) e}{1 - e^2} \right]^{-1},
\end{align}

and for the compressible model, we have
\begin{align}
    &\frac{de}{dt} = \nonumber \\ 
    & -\frac{\Omega^4(e) \beta}{\pi I_0} \left( \frac{\rho_{0}}{\rho^4} \right)^{1/3} (1 - e^2)^{1/3} \left[ 1 + \left( \frac{\rho_{0}}{\rho} \right)^{4/3} \frac{\Omega^2(e) \gamma_0}{\beta (1 - e^2)^{2/3}} \right] \times \nonumber \\ 
    & \bigg\{ g'(e) + \frac{g(e)}{3\Gamma - 4} \left[ \frac{4(\Gamma - 1)e}{1 - e^2} - \frac{A'_{3}}{A_{3}} \right] \bigg\}^{-1}.
\end{align}

\begin{figure}[!htbp]
	\includegraphics[width=1.1\columnwidth]{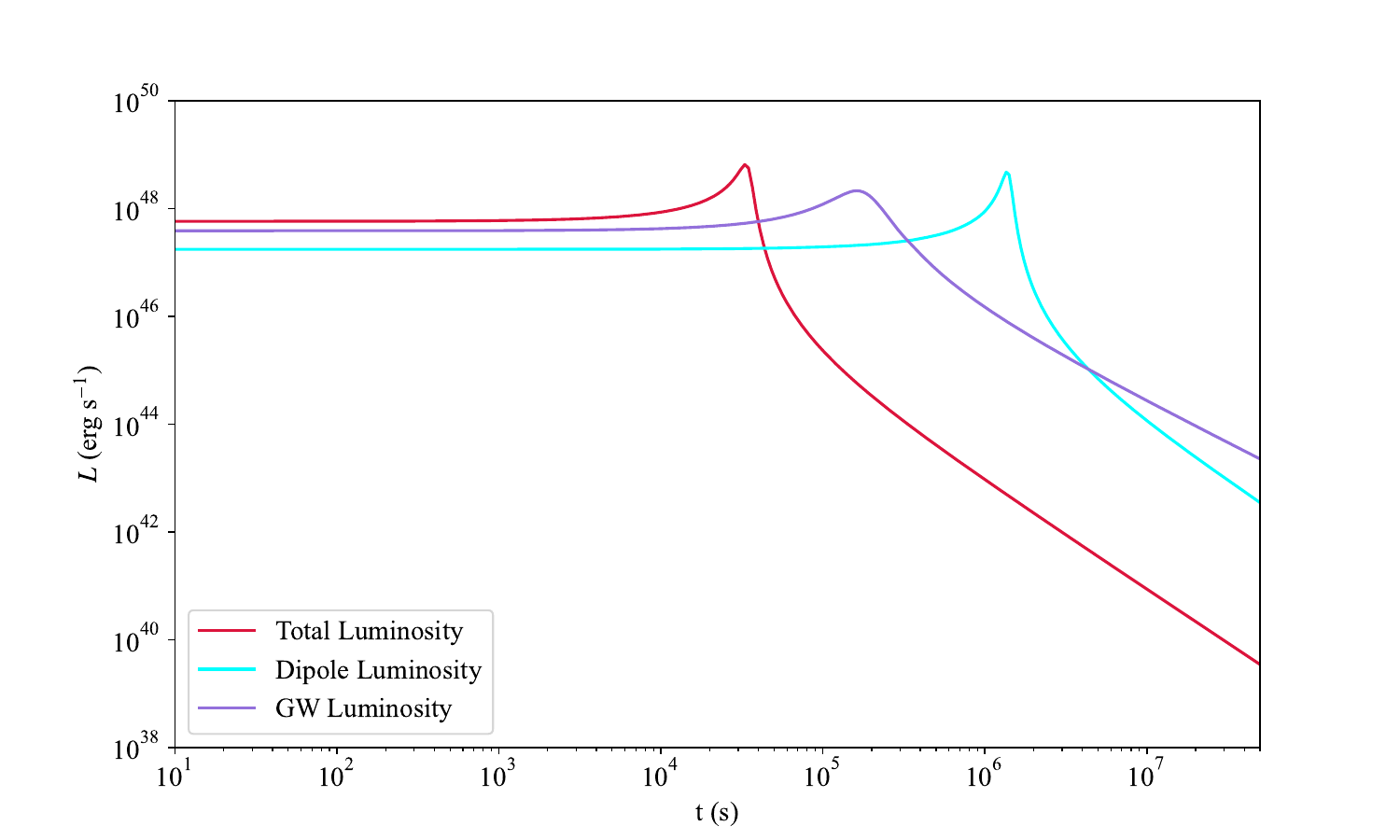}
    \caption{Comparing dipole, GW, and dipole+GW luminosity for the compressible model. Here, we assume that $M_{T}=2M\odot$, $B=5\times10^{15}G$ and $\epsilon=10^{-3}$.}
    \label{f14}
\end{figure}

Fig. \ref{f14} shows the resulting luminosity from EM dipole and GW radiations for incompressible model of a magnetar with $M_{T}=2M\odot$ and assuming $B=5\times10^{15}G$ and $\epsilon=10^{-3}$, the total luminosity (EM+GW) is also represented in this figure. According to this plot, it is seen that EM luminosity has a sharper peak, which appears with a delay after the GW peak time, and at the early times the GW luminosity, is dominated. For a time interval around dipole EM peak, it dominates and then decays with a faster rate than GW luminosity. The total luminosity, on the other hand, starts to decay earlier and faster with a slope larger than the dipole luminosity and less than GW one.

It is instructive to compare our results with those presented in \citet{Corsi:2009jt}, where it has been shown that the peak amplitude of the GW signal occurs with a delay compared to the GRB trigger. In our scenario, we see that for a given set of parameters, the GW flares can also even appear prior to EM ones. 
\subsection{Dipole and Quadrupole EM and GW Radiation}
Taking the quadrupole magnetic moment radiation into our consideration, the evolutionary equations for eccentricity in the incompressible and compressible models are given by
\begin{align}
     \frac{de}{dt} &=\frac{-\Omega^4(e) \beta}{\pi I_0 \rho} (1 - e^2)^{1/3}\times \\ \nonumber
    &  \bigg\{ 1 + \left[ \xi + \frac{\gamma_0}{(1 - e^2)^{2/3}} \right] \frac{\Omega^2(e)}{\beta} \bigg\} \left[ g' + \frac{4}{3} \frac{eg}{1 - e^2} \right]^{-1},
\end{align}

and
\begin{align}
     \frac{de}{dt} &= \\ \nonumber
    & \frac{-\Omega^4(e) \beta}{\pi I_0} (1 - e^2)^{1/3} \left( \frac{\rho_{0}}{\rho^4} \right)^{1/3} \times \\ \nonumber
    & \left[ 1 + \bigg\{ \xi + \gamma_0 \left( \frac{\rho_{0}}{\rho} \right)^{4/3} (1 - e^2)^{-2/3} \bigg\} \frac{\Omega^2(e)}{\beta} \right] \times \\ \nonumber
    & \bigg\{ g' + \frac{g}{3\Gamma - 4} \left[ \frac{4(\Gamma - 1)e}{1 - e^2} - \frac{A'_{3}}{A_{3}} \right] \bigg\}^{-1}.
\end{align}

\begin{figure}
\includegraphics[width=1.1\columnwidth]{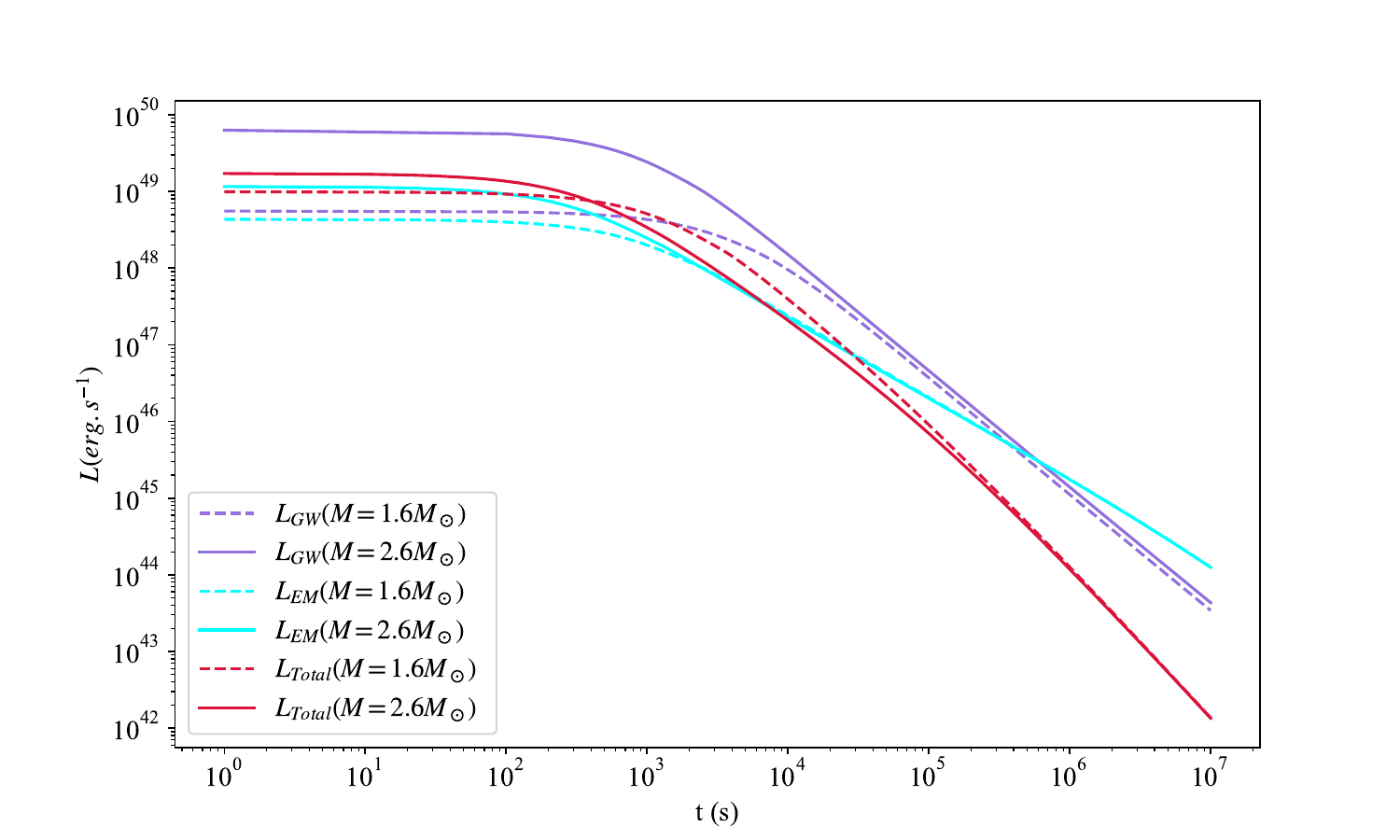}
    \caption{ GW, EM , and GW+EM luminosities for the incompressible model assuming $\epsilon=10^{-3}$ and $\kappa=100$ for total mass $M=1.6M\odot$ and $M =2M\odot$ as labeled.}
    \label{f15}
\end{figure}

\begin{figure}

	\includegraphics[width=1.1\columnwidth]{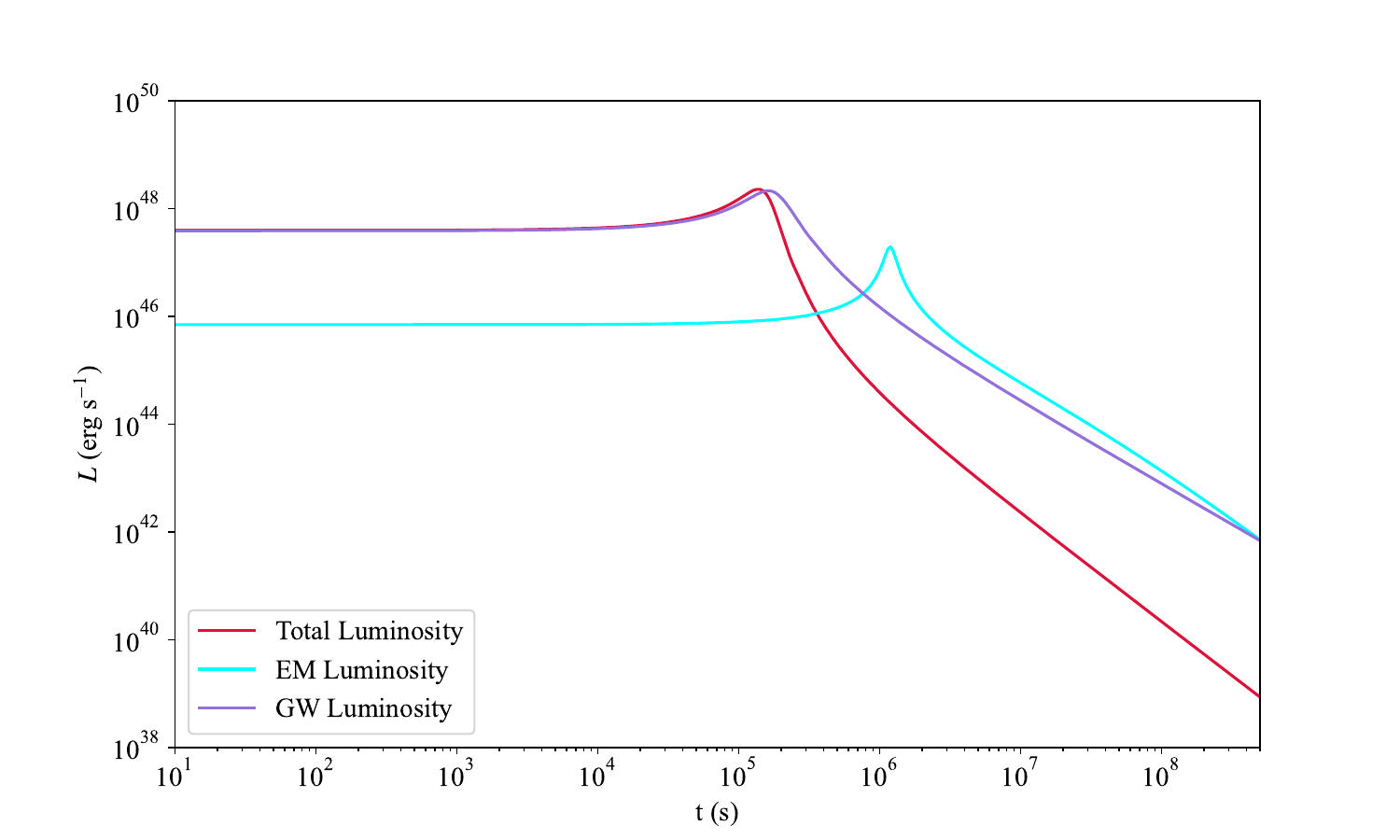}
 \includegraphics[width=1.1\columnwidth]{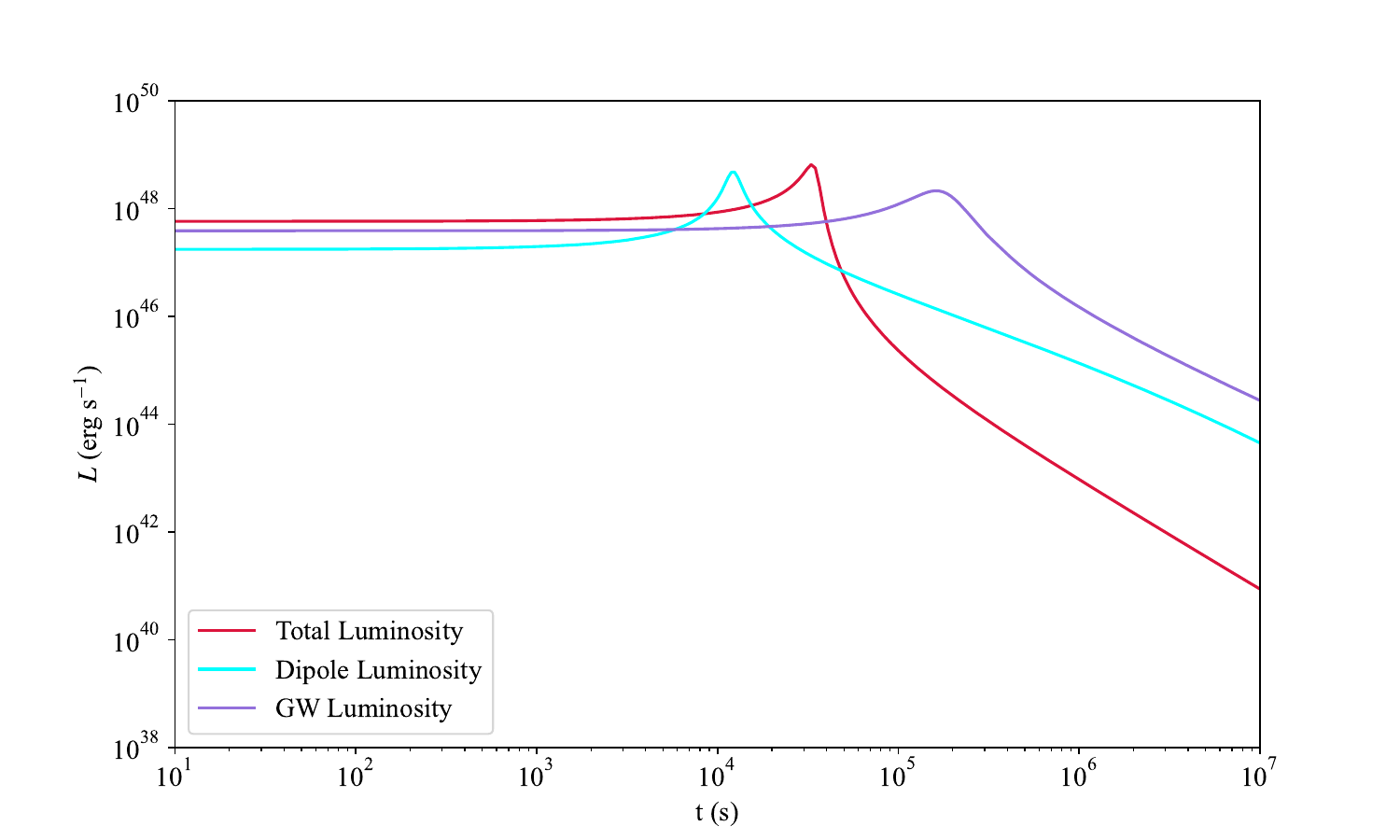}
    \caption{ Comparing EM (dipo+quad), GW, and total luminosity for a $2M\odot$ magnetar modeled with compressible fluid, applying  $\Gamma=1.43$ and $\epsilon=10^{-3}$, (top) assuming $B=10^{15}G$ and $\kappa=50$, 
(bottom) $B=5\times10^{15}G$, $\kappa=100$.}
    
    \label{f17}
\end{figure}

Fig. \ref{f15} illustrates a comparison of the EM, GW, and the total (GW+EM) luminosity for the incompressible model of nascent NSs with $M=1.6M_{\odot}$ and $M =2M_{\odot}$ regarding $\epsilon=10^{-3}$, $\kappa=100$ and $\Gamma=1.43$. We see that  the GW luminosity dominates at earlier times and EM radiation overcomes at later times. Similar to the behavior observed in Fig. \ref{f14}, the total radiation luminosity decays at a faster rate. 
In Fig. \ref{f17}, the behavior of EM, GW, and total luminosity are illustrated for two different sets of parameters. In both the top and bottom panels of Fig. \ref{f17}, a compressible magnetar with $\Gamma=1.43$ and $\epsilon=10^{-3}$ is assumed, but $B$ and $\kappa$ are different. This figure shows the crucial dependence of the  appearance time of  EM or GW peaks. In the top panel for  $B=10^{15}G$ and $\kappa=50$, the GW peak luminosity occurs prior to the EM peak luminosity, while in the bottom panel, the priority of these peaks are changed by assuming  $B=5\times10^{15}G$, $\kappa=100$. 
Our investigation in this section shows that in the case of the appearance of a peak within the luminosity, there is a delayed arrival time between peak in the GW and EM radiation, which provides a unique signature that can be confirmed by ongoing and upcoming multimessenger observations of high-energy astrophysics satellites and ground-based detectors.

 \begin{figure*}
\centering

	\includegraphics[width=.9\columnwidth]{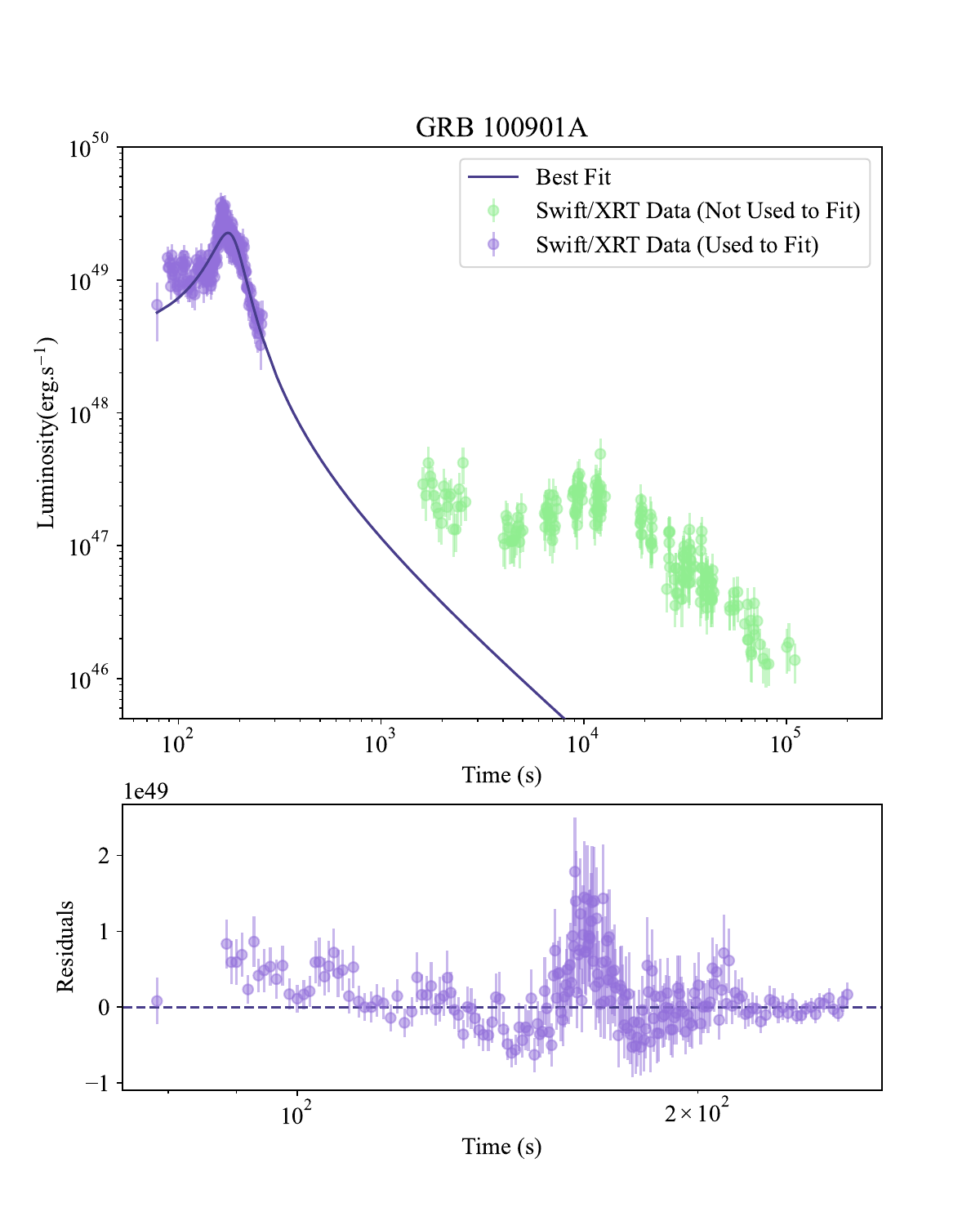}
 \includegraphics[width=1.1\columnwidth]{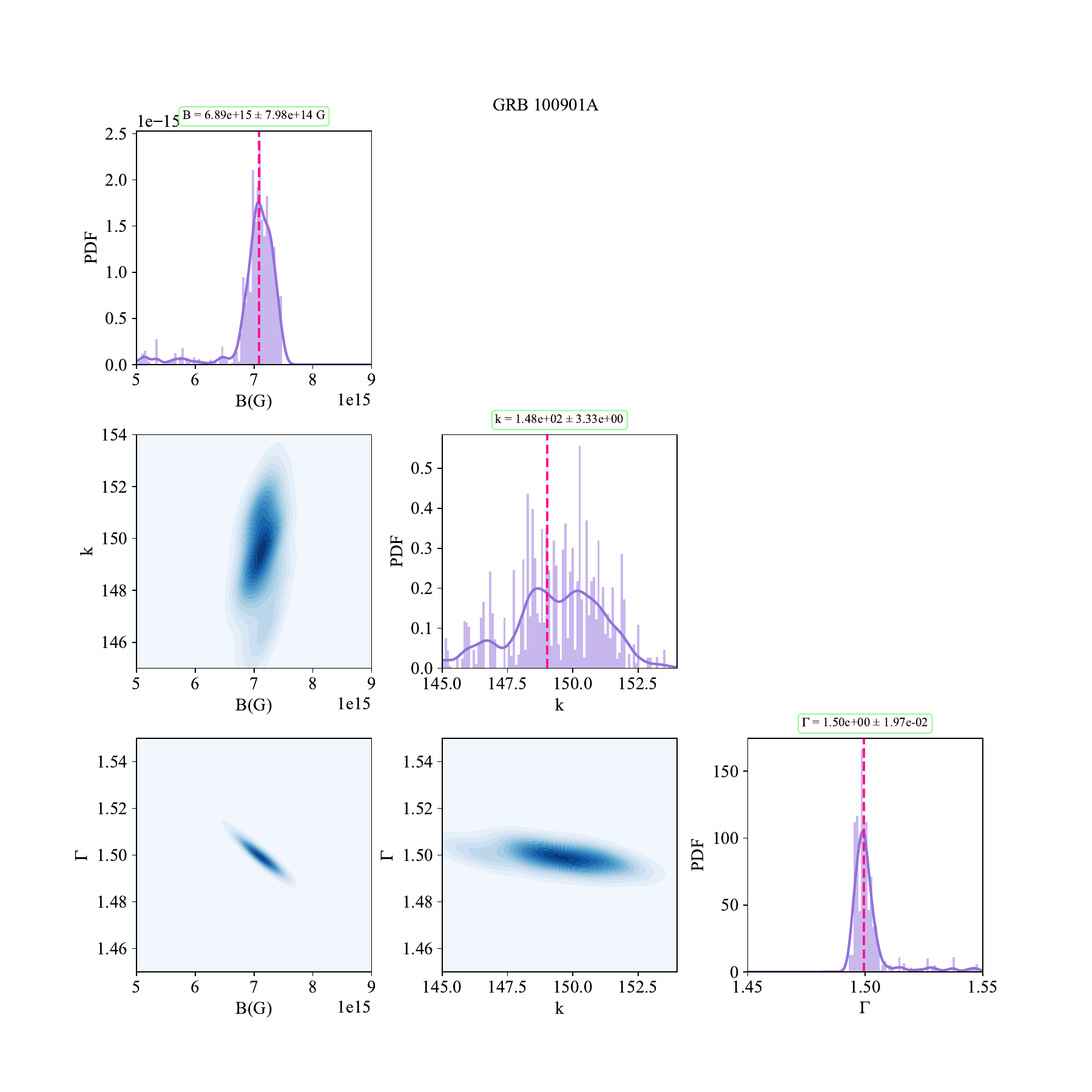}
    \caption{ In the left panel, the compressible model for a magnetar with $M=2M_{\odot}$ is used to fit the Swift-XRT data of GRB 100901A. The colored purple data  points (which cover the X-ray flares) are applied to find the best-fit model parameters $B_{\text{dip}}$, $\kappa$ and $\Gamma$. In the right panel, the posterior distributions for fitting parameters are plotted taking into account the histogram and estimated the PDF for each parameter.  The dashed magenta lines in PDF plots illustrate the median of the parameters, and the best-fit values are represented in the top of the diagrams.  }
    \label{d1}
    
\end{figure*}

 \begin{figure*}
\centering

	\includegraphics[width=0.9 \columnwidth]{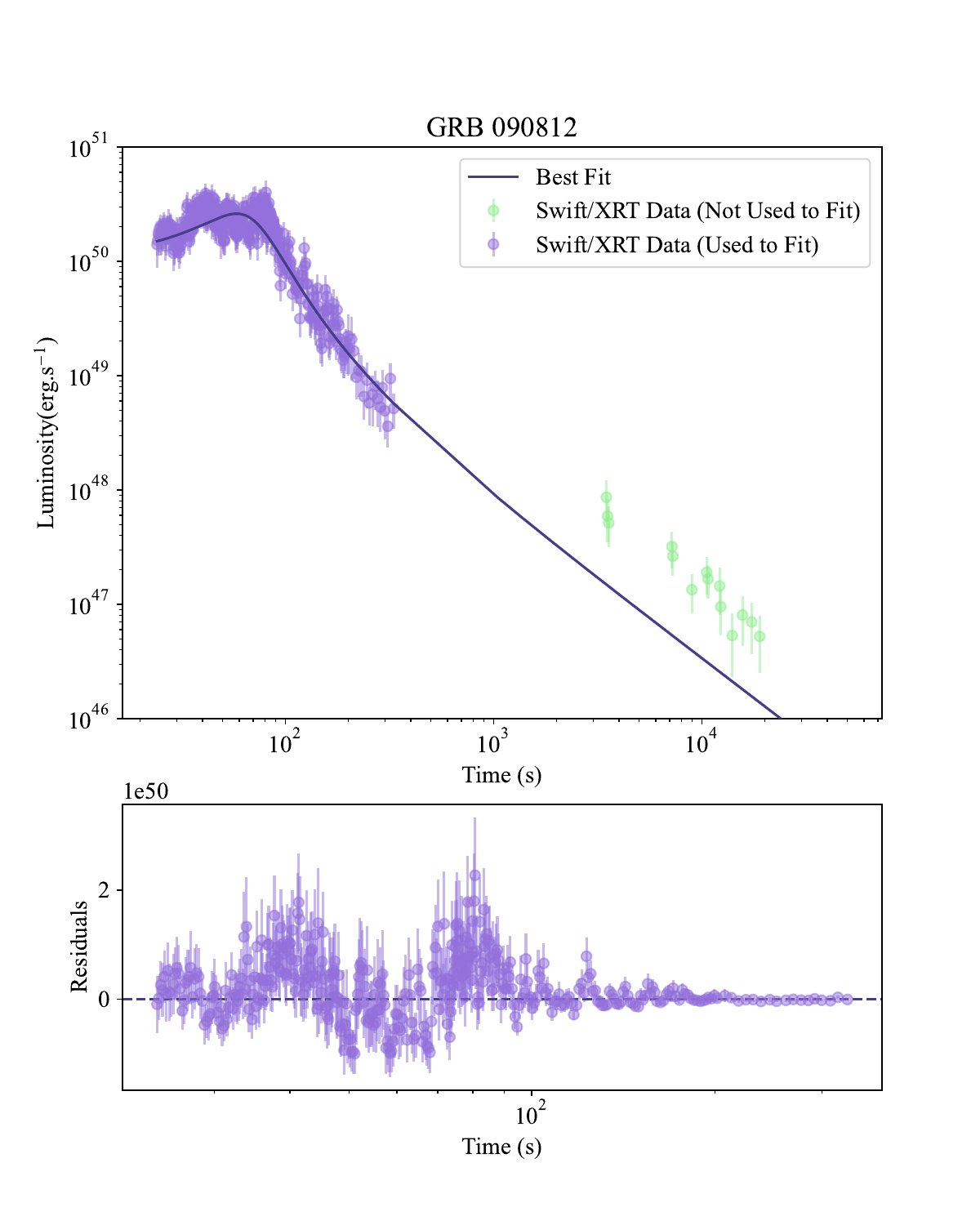}
 \includegraphics[width=1.1 \columnwidth]{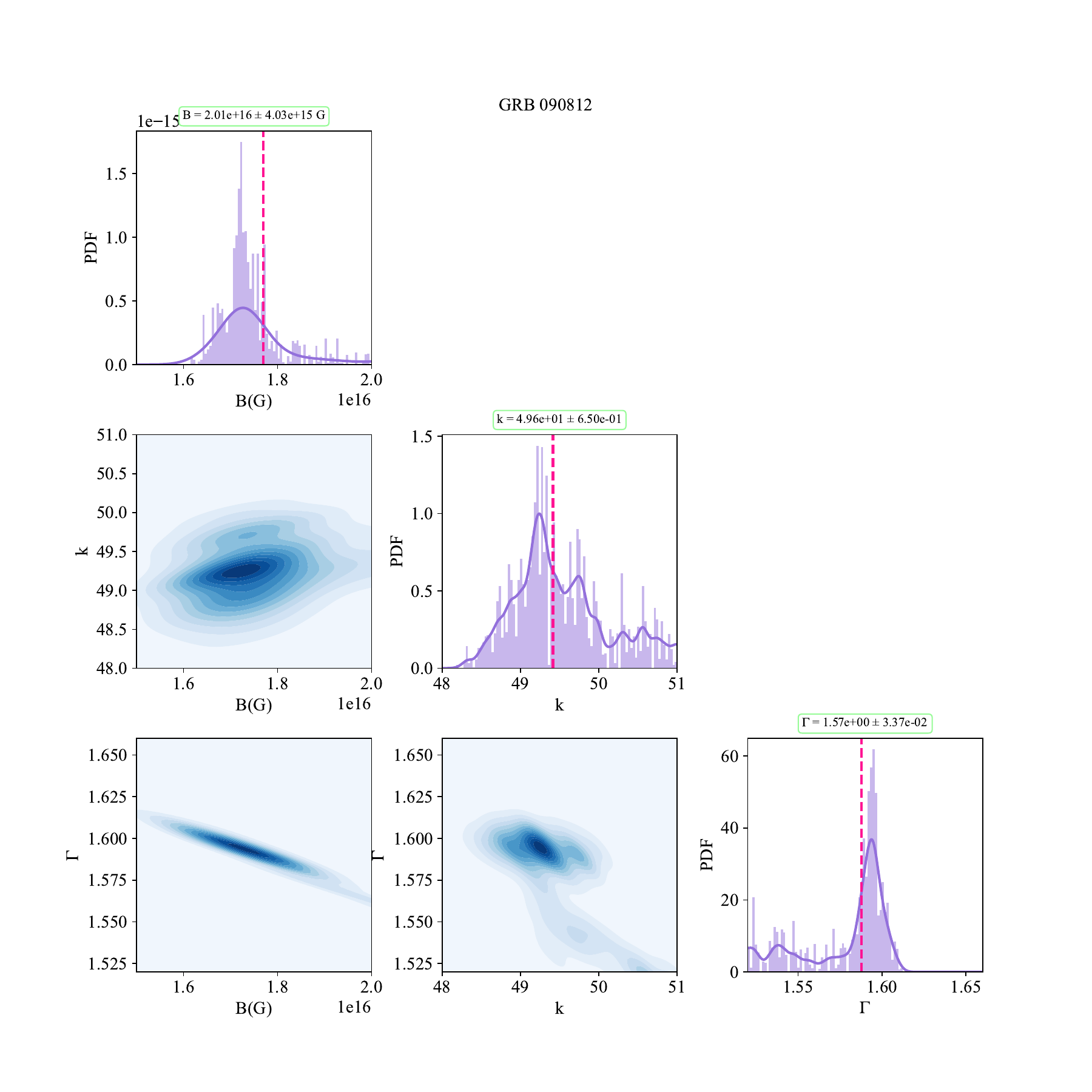}
    \caption{ The same as Fig. \ref{d1}, but for GRB 090812.}
    \label{d2}
\end{figure*}

 \begin{figure*}
\centering
	\includegraphics[width=0.9\columnwidth]{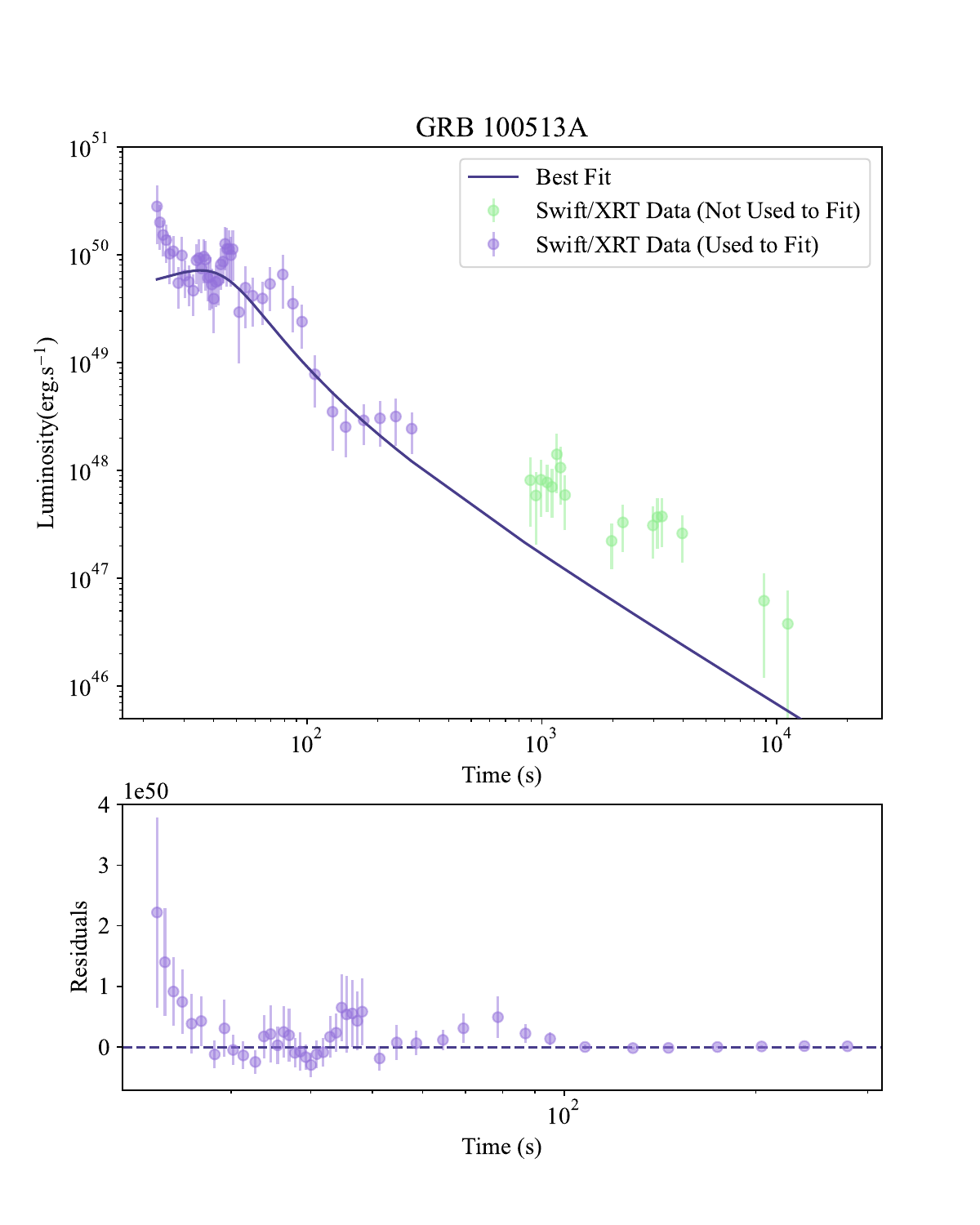}
\includegraphics[width=1.1\columnwidth]{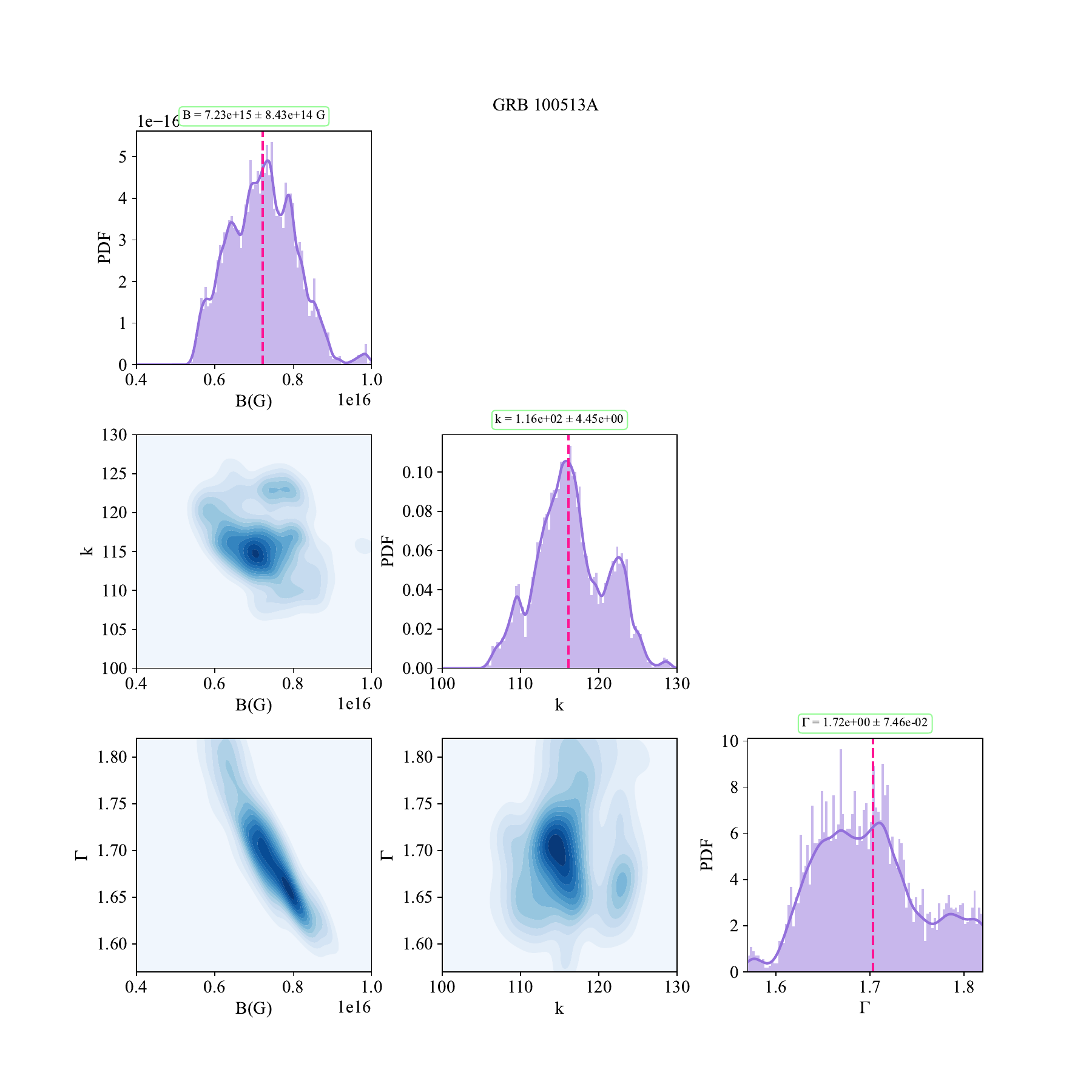}
 
    \caption{The same as Fig. \ref{d1}, but for GRB 100513A.}
    \label{d3}
\end{figure*}

 \begin{figure*}
\centering

	\includegraphics[width=0.9\columnwidth]{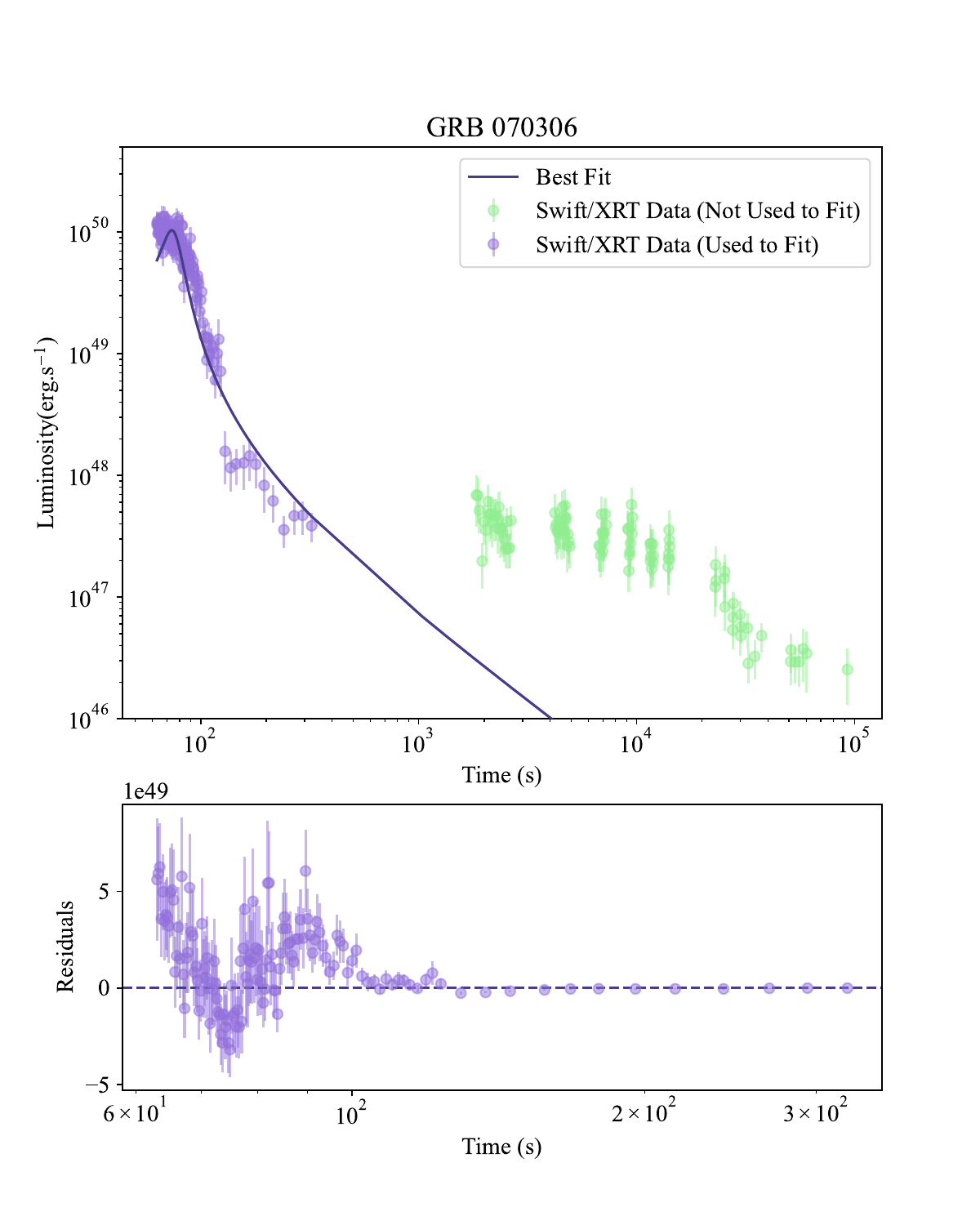}
 \includegraphics[width=1.1\columnwidth]{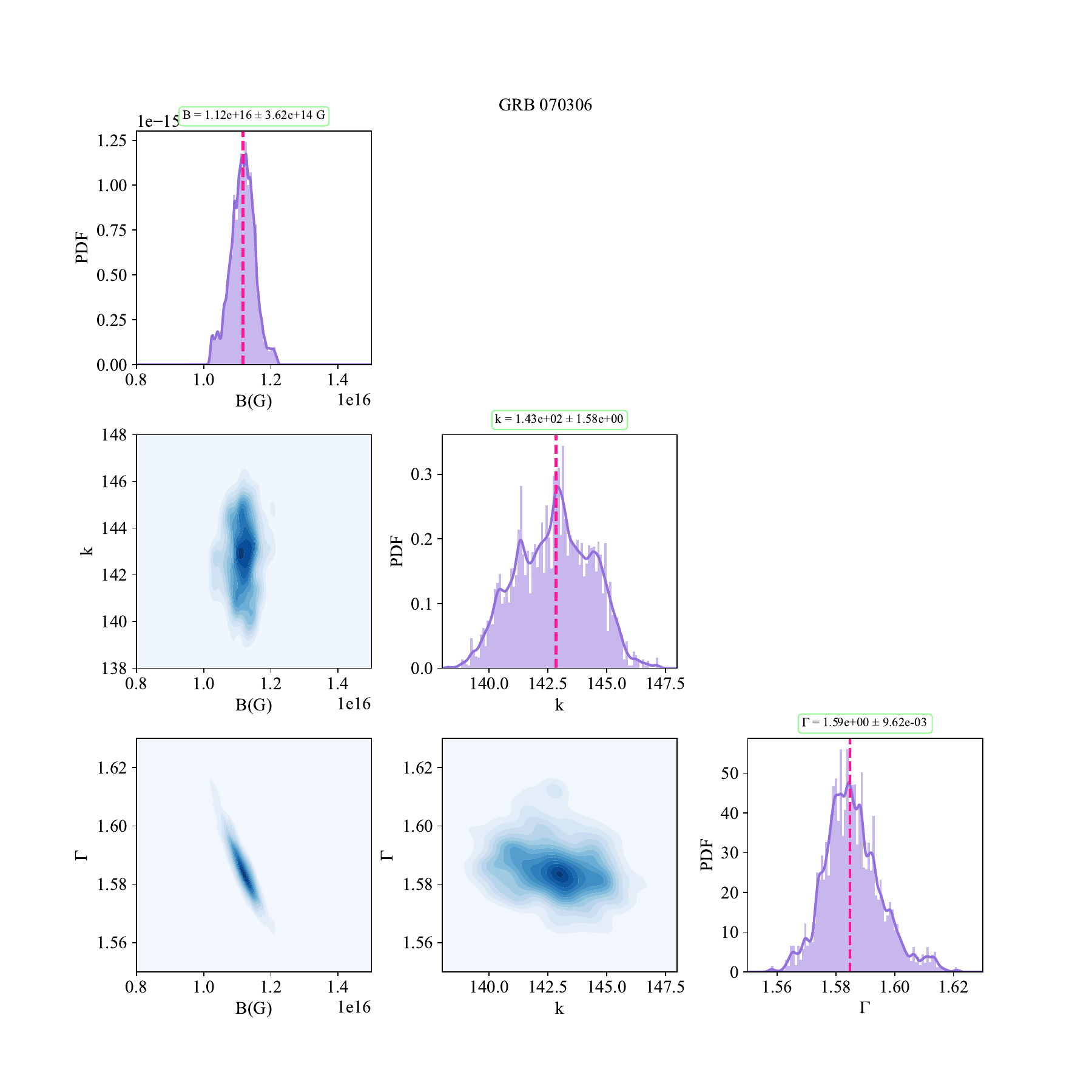}
    \caption{ The same as Fig. \ref{d1} but for GRB 070306.}
    \label{d4}
\end{figure*}

\section{Fitting GRB Flares within the compressible Magnetar Model} \label{fitt}
The GRB afterglow is observed in different energy bands from X-ray to the optical and radio bands, which follows a similar power law, but with significantly higher luminosity in the X-ray band. Consequently, the X-ray luminosity can serve as an indicator of the total (bolometric) afterglow luminosity. It is worth mentioning that due to the anisotropic radiation effects of GRBs introduced by the jet structure and also the instrumental limitations, the bolometric luminosity might be larger than the observed luminosity. Since the Swift-X-ray Telescope (XRT) generally offers the most comprehensive observations of afterglow evolution, in this paper, we rely on the afterglow luminosity data captured by Swift-XRT in order to compare with our model predictions. In some of the previous studies, the bolometric luminosity is estimated by multiplying  a factor of 2 - 7  to the
energy contained in the Swift-XRT band (0.3-10 keV) \citet{Rowlinson:2013ue,Rueda:2022mwh,Wang:2024fun}.

Here, we have focused on a GRB sample, including GRB 070306, GRB 090812, GRB 100513A and GRB 100901A, of which have known redshifts, and their modified light curves presented using a k-correction 
\citep{Bloom:2001ts}. All the GRB sources in our sample have the  maximum luminosity of the X-ray flares less than $10^{51} \text{erg s}^{-1}$ and the peak of the luminosity occurs in less than 200 s. In order to model the X-ray flares, we consider a new born NS with $2 M_{\odot}$ and free parameters are assumed to be the magnetic field $B$, the quadrupole-to-dipole ratio $\kappa$, and the adiabatic index $\Gamma$. Here, we adopt the Metropolis-Hasting (M-H) algorithm from the Markov Chain Monte Carlo (MCMC) method in Figs. \ref{d1} - \ref{d4} to find the best fit for the deformed newborn magnetar model (presented in sections \ref{se2} and \ref{se3}) taking into account the Swift-XRT data. The MCMC Python code is also employed in order to generate posterior  samples. We plotted a 2D distribution for  posteriors in the right panels of Figs. \ref{d1} - \ref{d4}  showing a pairwise correlation between parameters. In order to find the best-fit parameters, we used Swift-XRT data for X-flares occurring in less than a few hundred seconds (colored purple data points in  Figs. \ref{d1} - \ref{d4}) which covers the peak feature of the X-ray flares. For each parameter ($B$, $\kappa$ and $\Gamma$), we estimated the probability distribution function (PDF) with a Gaussian kernel density estimation (KDE) on the sample's histogram given in the right panels of Figs. \ref{d1} - \ref{d4}. The value for the width of Gaussian function is chosen to be 0.5, showing that the range of valid parameters to fit the data. 
The dashed magenta lines in the PDF plots indicate the median of each parameter sample. We provide $\chi^2_{\text{red}}=\chi^2/dof$ (chi-squared divided by the degrees of freedom) in Table. \ref{tab:grb_table}, and all the fitting parameters summarized in this table, with the errors are given as well. The model presented in this paper is limited to the stable Maclaurin spheroids; however, this investigation  can be extended to the Jacobi ellipsoid, in the case that  the eccentricity is larger than 0.81 and also for the case in which the three axes are different. Another direction to follow is considering spheroids in the presence of perturbations, taking into account the parameter space which is open to CFS instabilities, and to examine multimessenger signals in this regime.

We find that the peaks in Swift data can be produced by a softer EoS in which $\Gamma$ tends to $1.43$. Note that the plateau behavior in later times is given by  stiffer EoSs.  Generally, the magnetic field, $\kappa$ and EoS are not constant throughout the dynamical evolution of the deformed magnetars. These features should be taken into account in a more realistic
configuration and will be the subject of a future paper.
It is worth mentioning to note that while in the present paper, we focus on the modeling of GRB X-ray flares by  compressible new born magnetars, our model has also the potential to explain the X-ray plateau (in addition to X-ray flares) taking into account different types of the fluids  with multipolar structures of the magnetic field \citet{Wang:2024fun,10.1093/mnras/stac859}. We plan to perform a comprehensive statistical analysis for a large sample of X-ray flares in a future paper.
Taking into account the time dependency of the magnetic field and also including accretion materials onto a new born magnetar could be the origin of  the X-ray flares in GRB light curves and can be considered in a more generalized model.

\section{Conclusions}\label{sec6}

In this paper, we presented a scenario based on the dynamical evolution of a Maclaurin spheroid to model the GRB's central engine, taking into account both compressible and incompressible
fluids in order to explain different features of EM and GW  luminosities.  We evaluated the effect of the total mass $M_{T}$, the magnetic field $B$, the ratio of the magnetic quadrupole-to-dipole moment $\kappa$ and the adiabatic index $\Gamma$ on the light curves of a deformed NS. As it was presented in Sec. \ref{fitt}, our model was applied to fit X-ray flares in a GRB sample  observed by Swift-XRT with the peak luminosity  $\sim10^{49}$ - $10^{50}$ $\rm \text{erg}s^{-1}$ and the peak time less than $200$ s. The obtained values of the fitting parameters are $B\sim (6-20) \times 10^{15} G$, $\kappa\sim 49-150$ and $\Gamma\sim 1.5-1.7$. 
Moreover, considering a reasonable range of parameters, our model has the potential to explain X-ray flares with  the peak luminosity on the order of $10^{46}$ - $10^{51}$ $\rm \text{erg}/s$ which  occur within the wide range of $ 10$ to $10^4$ seconds after the Swift trigger time.

Our main motivation is inspired by \citet{SHAPIRO1990}, where the rotational evolution  of a  new NS originates from its deformation, which can result in both spin-down/spin-up  motion in various circumstances. We extended the analysis presented in \citet{SHAPIRO1990} to be used for the GRB central engine. In addition to performing the stability analysis over the parameter space, we introduced an additional quadrupole magnetic radiation term, which showed that the temporal spin-up of a newborn magnetar, for some specific EoSs, can lead to flaring features in the light curves. Previous studies primarily focused on explaining the X-ray plateau in the EM light curves of GRBs based on the rotational energy loss of magnetars \citet{Wang:2024fun,10.1093/mnras/stac859,Rueda:2022mwh,Lasky:2015olc}. However, in our work, we account for rotational distortion induced by the rapid rotation of NSs, which causes angular momentum to change due to the star's deformation. In our approach, the spin variation of a newborn magnetar is derived from the equilibrium equations of the Maclaurin spheroid, where the internal structure of the system through various EoSs affects the spin behavior of NSs.

In addition to analyzing the EM signal in our GRB model, we have examined the GW signal and its detectability based on the sensitivity range of GW detectors, as discussed in Sec. \ref{se4}. Using a reasonable set of parameters, the GW luminosity predicted by the model is $\sim10^{48}$ - $10^{52}$ $\rm \text{erg}/s$. Meanwhile, the peak time crucially depends on the model parameters and applied EoS, and the delay time between the peaks in GW and EM radiation (see Fig. (\ref{f17})) can serve as a unique signature for current and future multimessenger observations. By evaluating the characteristic GW strain for both spin-down regimes driven by EM and GW radiation, we have considered the detectability of the GW signals generated by a newly born deformed magnetar. A peculiar signature in the GW strain, tracing the spin evolution of the magnetar and its EoS, is also predicted in Fig. \ref{fig9}. The SNR for different GW detectors - ALIGO O4, ALIGO O5, KAGRA O5, ET, and CE - is estimated for a central frequency of 1 kHz in Table. \ref{t2}. This analysis shows that ET and CE have the highest detectability potential for magnetar GW signals.

In summary, through this study, we aimed to provide valuable insights into the underlying mechanisms driving the observed temporal behaviors in early-time GRB afterglows, such as the presence of X-ray flares, by considering a deformed magnetar as the GRB central engine. We demonstrated that the general properties of EM and GW luminosities crucially depend on the nature of the magnetar's EoS, which can generate flaring activities in both signals. Additionally, we predicted several distinct multimessenger features, highlighting the great potential of upcoming EM and GW missions to shed light on the nature of the GRB central engine.

\begin{table*}
\centering
\resizebox{1.9\columnwidth}{!}{%
\begin{tabular}{cccccccc} 
    \hline
    GRB Name & Redshift & $L_{\text{peak}}(\mathrm{erg\ s^{-1}})$ & $t_{\text{peak}}(\mathrm{s})$ & B (G) & $\Gamma$ & $\kappa$ & $\chi_{\text{red}}^2$ \\ [0.5ex] 
    \hline\hline
    GRB 070306 & 1.496 & $1.03\times10^{50}$ & 73.56 & $1.12^{+0.03}_{-0.03}\times10^{16}$ & $1.59^{+0.00}_{-0.00}$ & $142.8^{+1.58}_{-1.58}$ & 2.43 \\
    \hline
    GRB 090812 & 2.452 & $2.61\times10^{50}$ & 58.14 & $2.01^{+0.04}_{-0.04}\times10^{16}$ & $1.57^{+0.03}_{-0.03}$ & $49.6^{+0.65}_{-0.65}$ & 1.16 \\
    \hline
    GRB 100513A & 4.772 & $7.16\times10^{49}$ & 35.49 & $7.23^{+0.84}_{-0.84}\times10^{15}$ & $1.70^{+0.07}_{-0.07}$ & $116.1^{+4.45}_{-4.45}$ & 0.93 \\
    \hline
    GRB 100901A & 1.408 & $2.25\times10^{49}$ & 175.59 & $6.89^{+0.79}_{-0.79}\times10^{15}$ & $1.50^{+0.01}_{-0.01}$ & $148.0^{+3.33}_{-3.33}$ & 1.27 \\ 
    \hline
\end{tabular}
}
\caption{Constraints on the properties of the deformed newborn magnetar via fitting the X-ray light curve of GRBs. The table presents the fitting results, including the magnetic field strength (B), the adiabatic index $\Gamma$, and the ratio of the magnetic quadrupole-to-dipole moment ($\kappa$).}
\label{tab:grb_table}
\end{table*}

\section*{Acknowledgments}
We thank the anonymous referee for the valuable comments and suggestions. S.S and P.H would appreciate for valuable discussions during the conferences of High Energy Astrophysics and Cosmology in the era of all-sky in October 2024, Yerevan, Armenia. S.S and P.H are grateful to Massimo Della Valle for insightful comments and discussions.

\appendix

\section{\\Stability analysis and BH formation}\label{ap1}
\begin{figure*}[htbp]
    \centering
    \begin{minipage}{0.49\textwidth} 
        \centering
        \includegraphics[width=\textwidth]{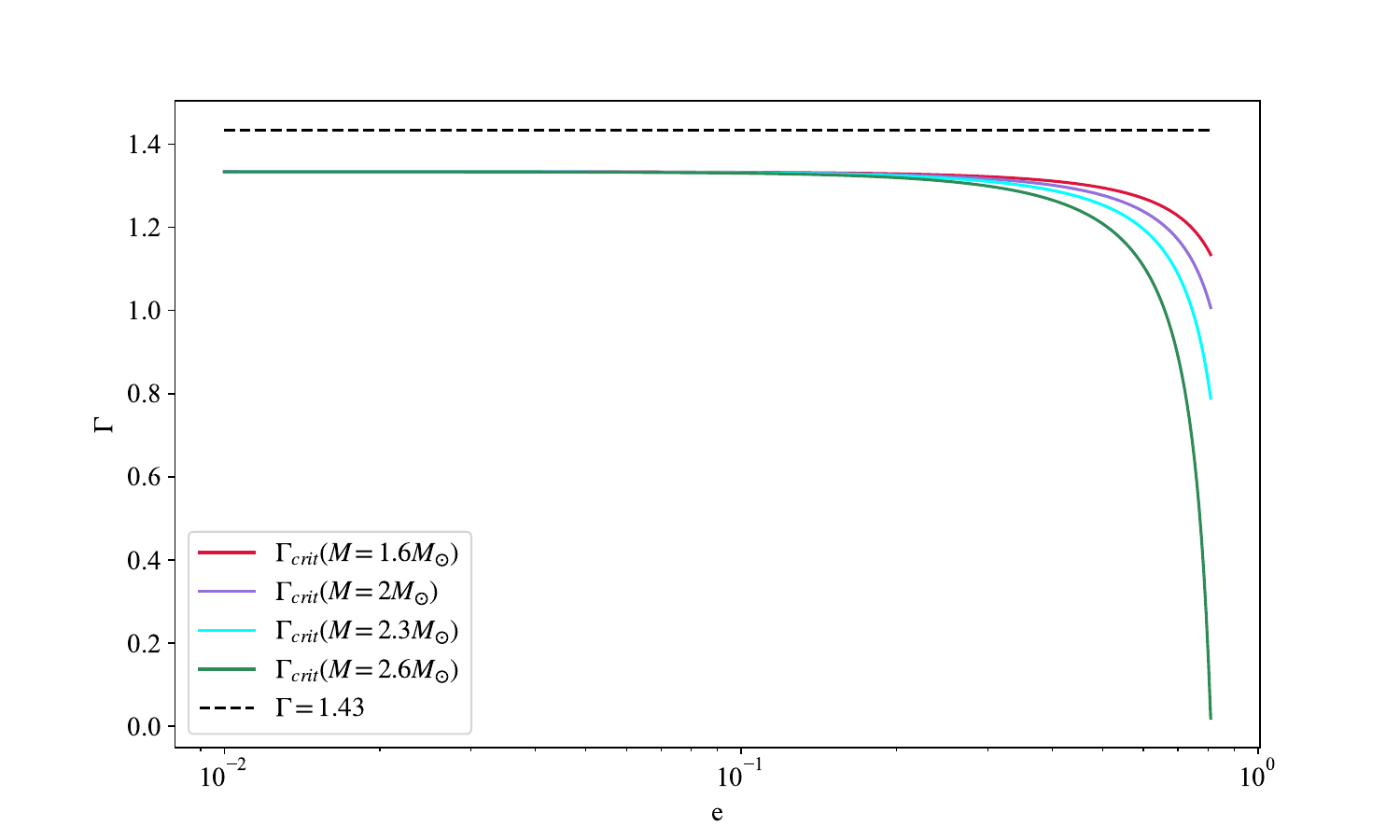}  
    \end{minipage}\hfill  
    \begin{minipage}{0.49\textwidth} 
        \centering
        \includegraphics[width=\textwidth]{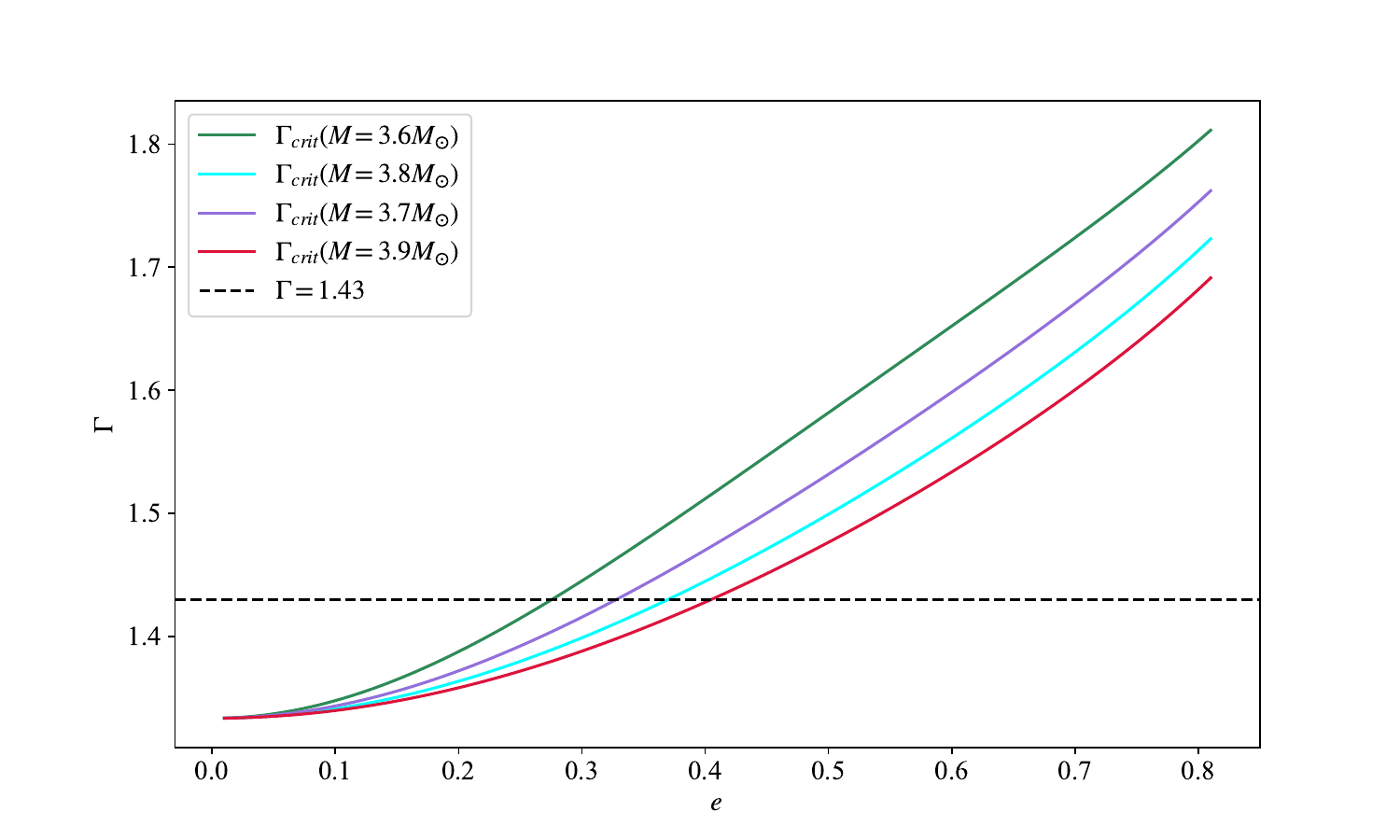}   
    \end{minipage}
    \caption{The plot shows the variation of the adiabatic index for various masses as labele. The colored lines show $\Gamma_{\text{crit}}$ for different masses, and the horizontal dashed lines display $\Gamma=1.43$.}
    \label{f9}
\end{figure*}

\begin{figure}
        \centering

	\includegraphics[width=0.7\columnwidth]{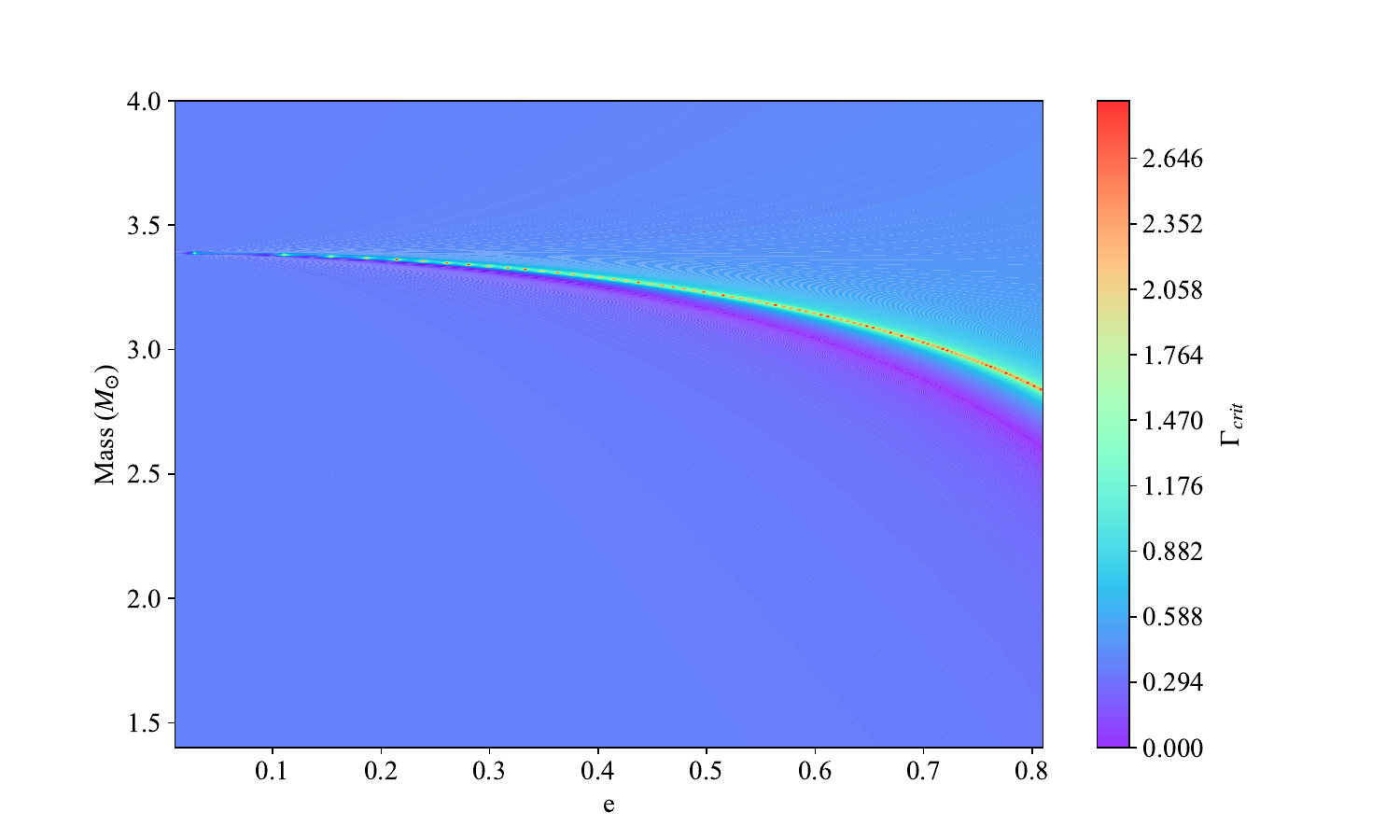}
    \caption{The density plot $\Gamma_{\text{crit}}$ is displayed as a function of both mass and eccentricity. This figure  contains essential information about the stability of NS, which is plotted in log scale.}
    \label{f13}
\end{figure}

In this appendix, we consider the criteria of the BH formation for both the incompressible and compressible models in order to determine  allowed region of the parameter space over the mass and adiabatic index. For the incompressible model, we should evaluate the Schwarzschild radius at $a_{0}=a(e=0)$ to investigate whether the BH is formed or not. But in the case of the compressible model, semi major axis $a$ is a more complicated function of the eccentricity, and it is not obvious that if $a_{0}$ gives the minimum radius.
Taking into account the logarithm of Eq. (\ref{eq:16}), we find that 
\begin{align}
   (\Gamma_{\text{crit}}-\frac{4}{3}) \log \left(\frac{\rho_{\text{crit}}}{\rho_{0}}\right)=\frac{3}{2}(1-e^2)^{2/3}A_{3}(e).
\end{align}
By considering the critical value for the density as below 
\begin{align}
   \rho_{\text{crit}}=\frac{M}{(4\pi/3)a_{\text{crit}}^3}=\frac{(1-e^2)^{1/2}}{(32\pi/3)M^2},
\end{align}
where $a_{\text{crit}}\sim R_{S}=2M$. Therefore, one can find a critical value for $\Gamma$ as (see  Eq. 63) in\citet{SHAPIRO1990}
\begin{align}\label{eq:42}
   \Gamma_{\text{crit}}=\frac{\log[2(1-e^2)^{-2/3}/3A_{3}(e)]}{\log[(32\pi/3)M^2\rho_{0}(1-e^2)^{-1/2}]}+\frac{4}{3}.
\end{align}
For an  NS with mass M, if $\Gamma<\Gamma_{\text{crit}}$ at a  given value of eccentricity, the star will collapse to a BH eventually. The stability analysis of the NS  are displayed in Figs. \ref{f9} and \ref{f13}.  In the left panel of Fig.  \ref{f9}, we present $\Gamma_{\text{crit}}$ for various masses in the range $(1.6-2.6)M_{\odot}$ as a function of eccentricity. For all chosen masses, all $\Gamma_{\text{crit}}$ curves lie below the $\Gamma=1.43$ line at the given range of eccentricity, meaning that an NS with a mass within this range remains stable during its evolution. On the right panel of Fig.  \ref{f9}, the mass range $(3.6-3.9)M_{\odot}$ is adopted. Within this range, all $\Gamma_{\text{crit}}$ curves are above $\Gamma=1.43$  line at the beginning of the evolution (for larger e), and they cross the  horizontal line at various eccentricity values. This  means that for a range of M and e, we are unable to have a stable configuration for $\Gamma=1.43$ and therefore, this part of the parameter space is excluded from our consideration.

In Fig. \ref{f13}, we present a density plot for $\Gamma_{\text{crit}}$ as a function of  mass and eccentricity. The presence of a singularity in the denominator of Eq. (\ref{eq:42}) for some values of  mass and eccentricity   leads to an infinite value for $\Gamma_{\text{crit}}$,  showing in the plot with the sharper color. Regarding the discussion mentioned above, this plot provides essential information about the maximum allowed value of  $\Gamma$ and therefore, the stable region of our parameter space.

\bibliography{sample631}{}
\bibliographystyle{aasjournal}

\end{document}